\documentclass[aps,pra,twocolumn,floatfix,superscriptaddress,nofootinbib]{revtex4} 

\usepackage{graphicx}
\usepackage{graphics}
\usepackage{amssymb}
\usepackage{amsmath}
\usepackage{latexsym}
\usepackage{color}
\usepackage{subfigure}
\usepackage{hyperref}

\hypersetup{
  colorlinks,
  citecolor=green,
  linkcolor=blue,
  urlcolor=blue}

\newcommand{\tr}{{\rm tr \thinspace}}
\newcommand{\bra}[1]{\left\langle{#1}\right\vert}
\newcommand{\ket}[1]{\left\vert{#1}\right\rangle}

\newcommand{\expect}[1]{\langle{#1}\rangle}

\newcommand{\beq}{\begin{equation}}
\newcommand{\eeq}{\end{equation}}
\newcommand{\bqa}{\begin{eqnarray}}
\newcommand{\eqa}{\end{eqnarray}}
\newcommand{\nn}{\nonumber}

\newcommand{\erf}[1]{Eq.~(\ref{#1})}
\newcommand{\erfs}[2]{Eqs.~(\ref{#1})-(\ref{#2})}
\newcommand{\dg}{^\dagger}

\newcommand{\ddt}{\frac{\text{d}}{\text{d}t}}
\newcommand{\Op}[1]{\boldsymbol{\mathsf{#1}}}

\newcommand{\eg}{\emph{e.g.},~}
\newcommand{\ie}{\emph{i.e.},~}

\def\etal{\eta_{l}}
\def\etam{\eta_{\rm m}}

\def\be{\begin{equation}}
\def\ee{\end{equation}}
\def\bea{\begin{eqnarray}}
\def\eea{\end{eqnarray}}

\newcommand{\BQIC}{Berkeley Center for Quantum Information and Computation, Berkeley, California 94720 USA}
\newcommand{\DeptChem}{Department of Chemistry, University of California, Berkeley, California 94720 USA}
\newcommand{\DeptPhys}{Department of Physics, University of California, Berkeley, California 94720 USA}
\newcommand{\Sandia}{Digital \& Quantum Information Systems, Sandia National Laboratories, Livermore, CA 94550 USA}

\begin{document}

\title{Continuous joint measurement and entanglement of qubits in remote cavities}
\date{\today}

\author{Felix Motzoi }
\affiliation{\BQIC}
\affiliation{\DeptChem}
\affiliation{\DeptPhys}

\author{K. Birgitta Whaley}
\affiliation{\BQIC}
\affiliation{\DeptChem}

\author{Mohan~Sarovar}
\email{mnsarov@sandia.gov}
\affiliation{\Sandia}

\begin{abstract}
We present a first principles theoretical analysis of the entanglement of two superconducting qubits in spatially separated microwave cavities by a sequential (cascaded) probe of the two cavities with a coherent mode, that provides a full characterization of both the continuous
 measurement induced dynamics and the entanglement generation. 
We use the SLH formalism to derive the full quantum master equation for the coupled qubits and cavities system, within the rotating wave and dispersive approximations, and conditioned equations for the cavity fields.  We then develop effective stochastic master equations for the dynamics of the qubit system in both a polaronic reference frame and a reduced representation within the laboratory frame.  We compare simulations with and analyze tradeoffs between these two representations, including the onset of a non-Markovian regime for simulations in the reduced representation.
We provide conditions for ensuring persistence of entanglement and show that using shaped pulses enables these conditions to be met at all times under general experimental conditions. The resulting entanglement is shown to be robust with respect to measurement imperfections and loss channels. We also study the effects of qubit driving and relaxation dynamics during a weak measurement, as a prelude to modeling measurement-based feedback control in this cascaded system.

\end{abstract}


\maketitle

\section{Introduction}
Entanglement between remote parties is a key resource in many quantum information applications, including quantum teleportation, quantum key distribution, and quantum metrology, and is also central to the notion of a \emph{quantum network} \cite{Kim-2008}. Distributed entangled states are also a critical component for scalable quantum computing, since they enable long-range gates between spatially separated qubits \cite{Gottesman:1999kw}. Accordingly many different approaches have been proposed to distribute or generate entangled states among systems that are significantly spatially separated. Distributing entangled states after preparation at a central location is practically challenging since decoherence in distribution channels typically degrades entanglement, \eg \cite{Aspelmeyer:2003jy, Peng:2005de}. Alternatively, a long-range coupling between remote systems can be 
engineered by exchanging single quanta, and entanglement can be generated this way, as has been recently demonstrated for atoms, photons, and combinations thereof \cite{Lettner-2011, Ritter:2012jn}.

A fundamentally distinct approach for preparing entangled states of systems residing at remote locations is to perform a \emph{joint measurement} on them. Most proposals for achieving such joint-measurement-enabled entanglement interfere photons that are spontaneously emitted by atoms (or artificial atoms) in such a way that subsequent detection of a photon makes the identity of the emitter indiscernible, \eg \cite{Cabrillo-1999,Bose-1999,Duan:518484,Feng-2003,Simon-2003,Browne-2003}, and thus projects the remote atoms into an entangled state. The degree of entanglement generated is heavily dependent on both the quality and stability of the interferometer and efficiency of detection of spontaneously emitted photons. As a result achieving high fidelity entangled states with this approach is challenging, although several proof-of-principle experiments have demonstrated validity of the approach \cite{Chou:2005kx, Moehring:2007bu}. An alternate approach is to perform a joint measurement by sequentially interacting two systems with a coherent light mode. This has been explored as a method for generating entanglement theoretically \cite{Duan-2000, Cla.Pen.etal-2003} and experimentally implemented using collective excitations of atomic clouds \cite{Julsgaard:2001tu}. Most recently, a sequential probe has been utilized to probabilistically entangle superconducting qubits in separate microwave cavities \cite{Roch:2014ey}. In this work we develop a theoretical description of that experiment from first-principles, providing a rigorous and general theoretical framework for the generation of entanglement by joint dispersive measurement of qubits in distinct cavities and analyzing in detail the potential and limitations of entanglement generation in this setting. 

\begin{figure}
\centering \includegraphics[width=1.01\columnwidth]{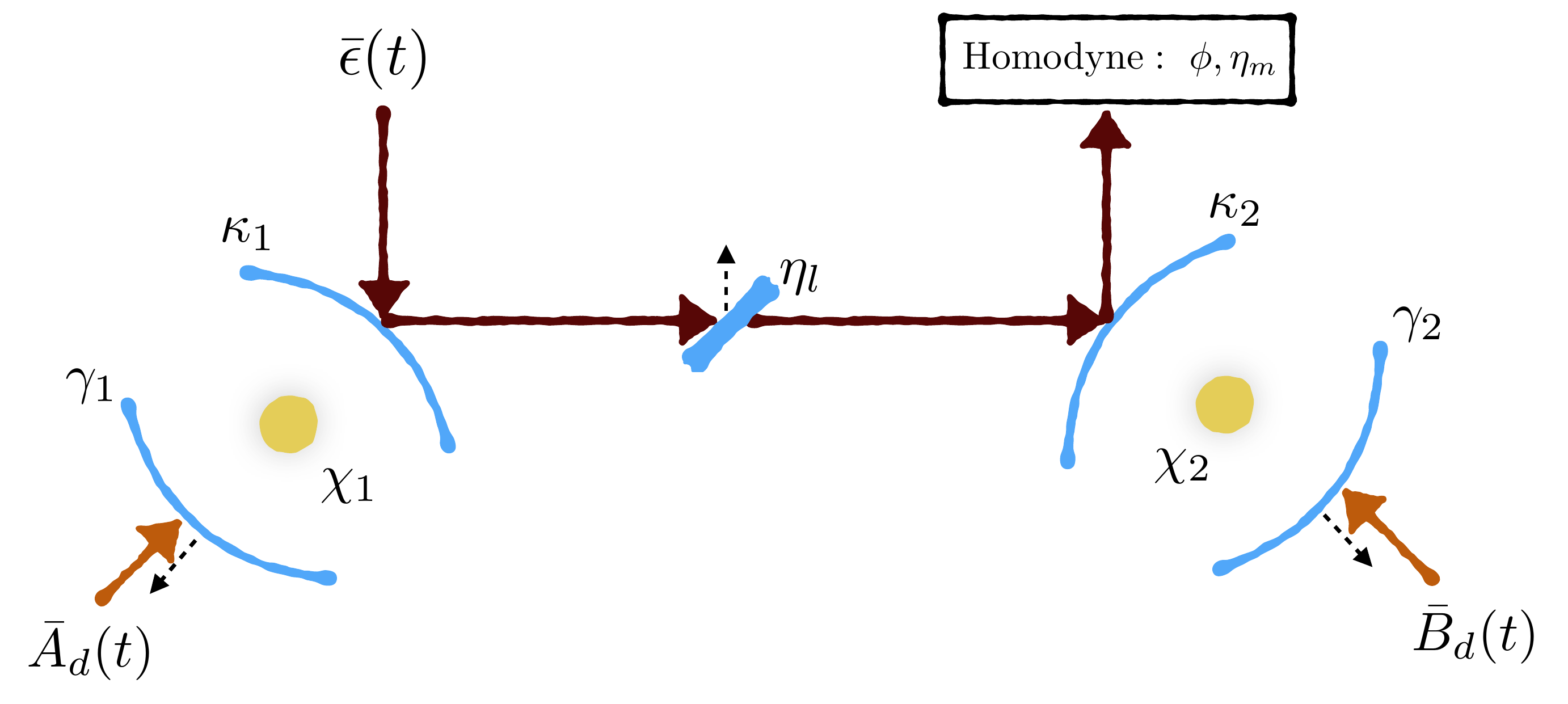}
\caption{Sequential probe of two spatially separated cavities,
each containing a qubit coupled dispersively with strength $\chi_i$ to its fundamental cavity mode. The cavities are asymmetric with the $\kappa$ ports being more transmissive than the $\gamma$ ports. The beam-splitter between the cavities models the losses induced by the circulator that enforces one-way field propagation between cavities. We allow for arbitrary coherent state drives ($\bar{A}_d(t)$ and $\bar{B}_d(t)$) into the weakly coupled ports of both cavities. The output field that results after sequential reflection of the probe field from both cavities, $z(t)$, is measured by a homodyne detector with efficiency $\eta_{\rm m}$ and at a phase $\phi$ with respect to the probe $\bar{\epsilon}(t)$.
}
\label{fig:apparatus}
\end{figure}

In the dispersive interaction regime, a coherent mode reflected off a cavity with an embedded qubit acquires a phase shift that depends on the internal state of the qubit. 
This motivates the essential idea behind the entanglement generation scheme we study here, namely, to perform a 
measurement of the parity of the qubit pair excitation state by sequentially probing the two cavities that contain them, and performing a homodyne measurement of the total phase acquired by the twice-reflected probe field.  Fig. \ref{fig:apparatus} shows a schematic of the apparatus. Ideally, the qubit observable that corresponds to this measurement, which we shall refer to as a half-parity measurement, takes the form 
$O_{hp} = \Op\sigma_z^1 + \Op\sigma_z^2$, where $\Op\sigma_z^i$ is the Pauli-$Z$ operator on the $i^{th}$ qubit. This observable cannot distinguish between the qubit basis states $\ket{01}$ and $\ket{10}$ \footnote{Here, and in the following, for conciseness we omit tensor products when writing multiparty states.}. Therefore if the initial state of the two qubits is the 
equal superposition state $\ket{\Psi_0} = \frac{1}{2}(\ket{00}+\ket{01}+\ket{10}+\ket{11})$, the ideal half-parity measurement will yield the states $\ket{00}$ or $\ket{11}$, each with a probability $1/4$, 
or the state $\frac{1}{\sqrt{2}}(\ket{01}+\ket{10})$
 with probability $1/2$.  
This is to be distinguished from the full parity measurement of $O_{fp} = \Op\sigma_z^1  \Op\sigma_z^2$, which yields the states $\frac{1}{\sqrt{2}}(\ket{00}+\ket{11})$ and $\frac{1}{\sqrt{2}}(\ket{01}+\ket{10})$ with equal probabilities.  In the following we develop a detailed model of this sequential probe measurement from first principles, including all non-idealities present in the experiment of Ref. \cite{Roch:2014ey}. Although we develop the model within the context of the superconducting qubit experiment of Ref. \cite{Roch:2014ey}, it applies more generally to any implementation of cavity-QED, including in the optical domain.

The remainder of the paper is structured as follows. Section \ref{sec:full_model} presents the dynamical model for full system derived from cascaded systems theory \cite{Gar-1993,Car-1993,Gough:2012fl}. Then in section \ref{sec:polaron} we perform a two-cavity polaron transformation on the model to obtain an exact, dressed description for the qubits alone that is easier to simulate than the full dynamical model. 
This polaron transform and subsequent derivation of an effective master equation for the dressed qubits constitutes a generalization of the methods first presented in Ref. \cite{Gam.Bla.etal-2008}. 
In Sections \ref{sec:reduced_lab} and \ref{sec:red_sme} we simulate the qubit-reduced dynamics in the laboratory frame, including the loss and revival of coherence between qubits.   Section \ref{sec:entanglementcondition} derives physical requirements and criteria for generating entanglement between the remote qubits.  Section \ref{sec:results} provides simulation data for a range of realistic experimental parameters and discusses the viability of obtaining high-grade concurrence betwen the qubits. Section \ref{sec:CW} develops a perturbative treatment of qubit driving during the continuous measurement. Finally, section \ref{sec:disc} provides a summary and assessment of the benefits and possible extensions of this approach for other quantum processing tasks with superconducting qubits.  

\section{Derivation of theoretical model}
\label{sec:theory}
\subsection{Full model of cascaded cavities}
\label{sec:full_model}
Consider the apparatus shown in Fig. \ref{fig:apparatus}. 
Each cavity has two ports, with asymmetric transmittivities. The ``input" port on each cavity is low transmittivity ($\gamma_i$) and the ``output" port is high transmittivity ($\kappa_i$). The probe field $\bar \epsilon(t)$ interfaces with the output ports of both cavities a distance $L$ apart (in \cite{Roch:2014ey} a distance $L=1.3$m was achieved), before impinging on the homodyne detector. The cavities are operated in the dispersive regime, where the Hamiltonians in the two cavities are given by 
\bqa
H_A&=& \Delta_{1} \Op a\dg \Op a +  \chi_1 \Op a\dg\Op a \Op \sigma_z^1, \nn\\
H_B &=&  \Delta_{2} \Op b\dg \Op b  + \chi_2 \Op b\dg\Op b \Op \sigma_z^2, \label{eq:dispersive}
\eqa
respectively, where $\Op a (\Op b)$ is the annihilation operator for the fundamental mode in cavity $1 (2)$, $\Op \sigma_z^{1(2)}$ is the Pauli $z$ operator for qubit 1(2), $\Delta_i \equiv \omega_d - \omega_r^i$ is the detuning of cavity $i$ from the probe field $\bar \epsilon(t)$ (frequency $\omega_d$), and $\chi_i$ is the qubit-cavity coupling in the dispersive regime. These Hamiltonians are in the interaction frame with respect to the free Hamiltonians for the qubits: $-\frac{\omega_1}{2} \Op \sigma_z^1 -\frac{\omega_2}{2} \Op \sigma_z^2$. The input ports for the two cavities can also be used for driving the cavity or qubits at their respective frequencies (with fields $\bar A_d(t)$ and $\bar B_d(t)$, respectively), for state initialization and tomography. We will see that the coherent drive $\bar B_d(t)$ will be useful for compensating against asymmetries in the parameters between the two cavities and qubits. To minimize back-reflection of the probe field and ensure its unidirectionality, a circulator is inserted between the two cavities. Losses associated to this circulator will be included in the model developed below.
The parameters of the system, including cavity transmitivities, losses, cavity coupling, are labeled in Fig.  \ref{fig:apparatus}.

\begin{figure}
\subfigure[]{\includegraphics[width=0.95\columnwidth]{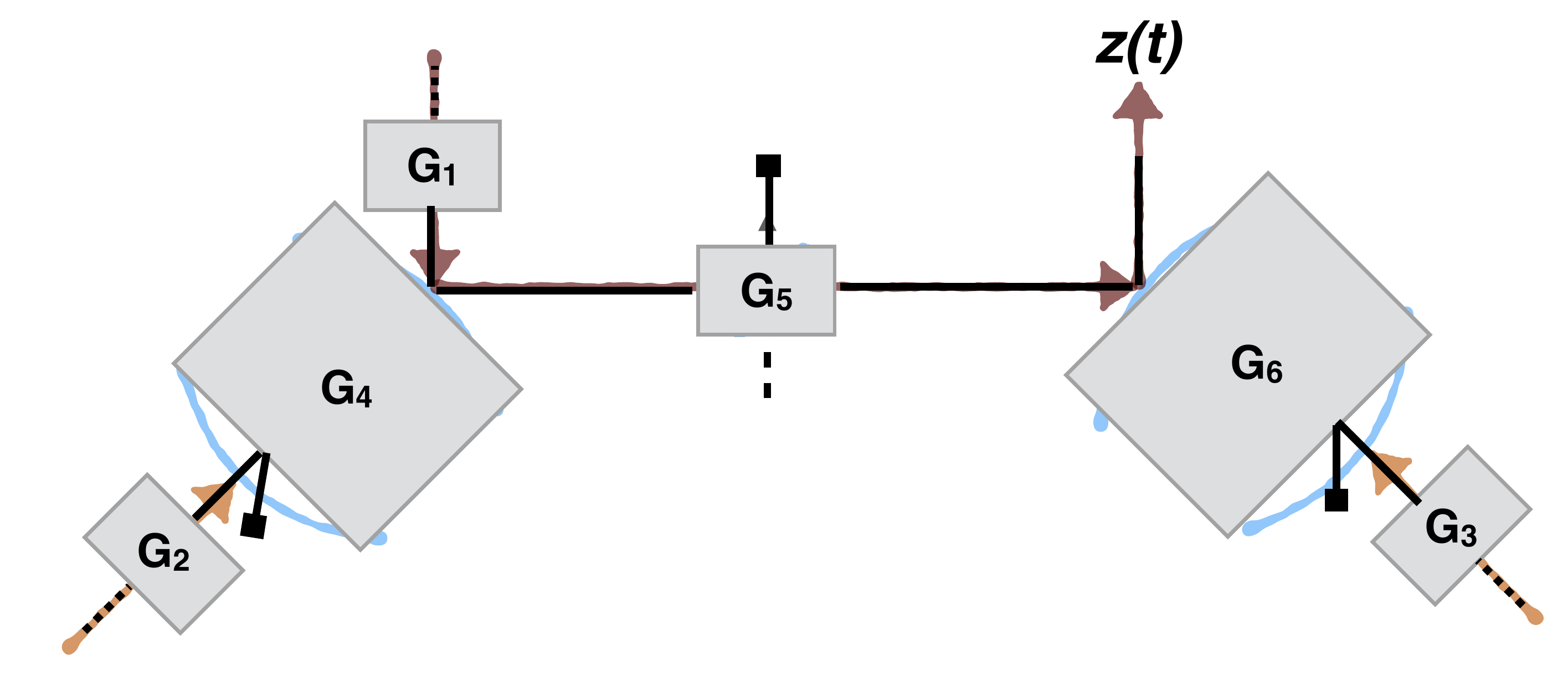}}
\subfigure[]{\includegraphics[width=1\columnwidth]{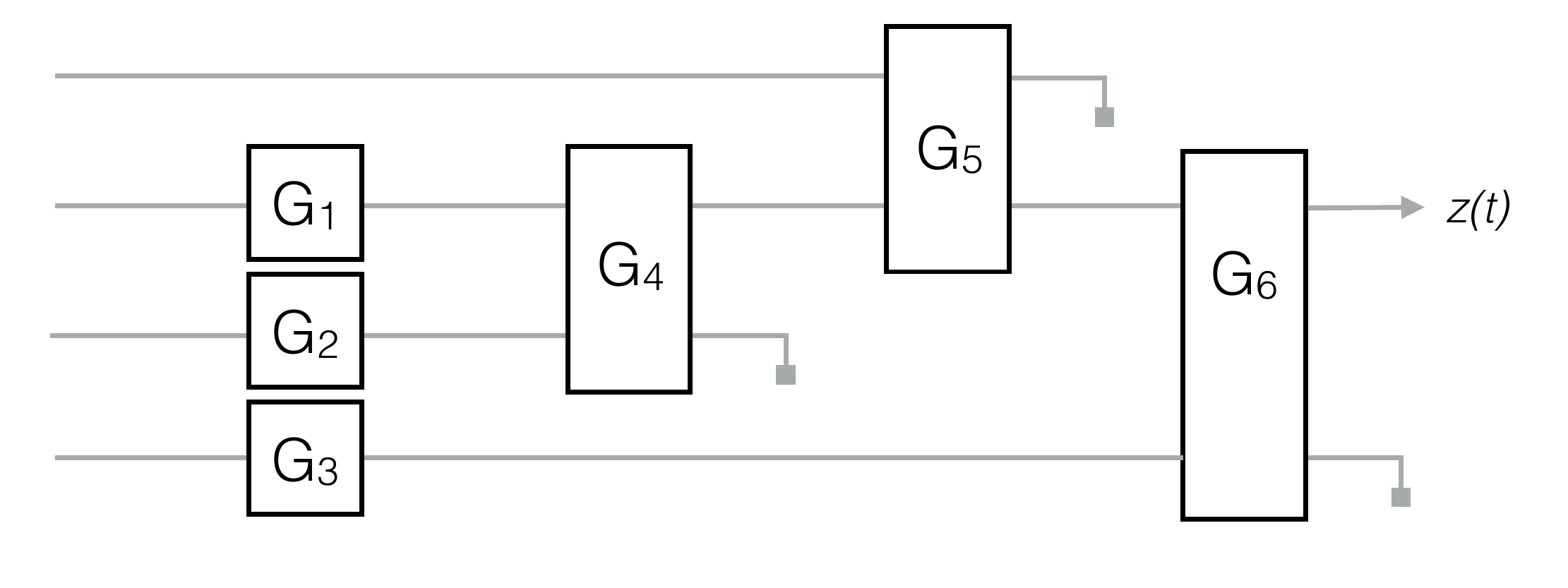}}
\caption{SLH network decomposition of the apparatus in Fig. \ref{fig:apparatus}.
The SLH triples $(S, L, H)$ for each block are specified in the main text. (a) shows an SLH block representation superimposed on the experimental apparatus, and (b) shows the SLH block diagram redrawn more conventionally with inputs on the left and outputs on the right. The explicit forms for the resulting $S$ (scattering), $H$ (self-energy), and $L$ (coupling) matrices for the overall model are given in the Appendix. All inputs are in the vacuum state (indicated by dotted lines in (a)) and the single monitored output is $z(t)$, the field reflected from the output port of cavity 2. All other outputs are not monitored: this is indicated in (a) and (b) by their termination. \label{fig:apparatus_slh}}
\end{figure}

The dynamical model for the apparatus described in Fig. \ref{fig:apparatus} can be derived using the cascaded cavity theory of Gardiner and Carmichael \cite{Gar-1993,Car-1993}, or by the modern SLH quantum network theory \cite{Gou.Jam-2009,Gough:2012fl}. We utilize the latter here and Fig. \ref{fig:apparatus_slh} presents an SLH network diagram that is equivalent to the apparatus in Fig. \ref{fig:apparatus}. Each block $G_i$ is specified by an $(S,L,H)$ triple, where $S$ is a scattering matrix, $L$ a coupling matrix and $H$ a self-energy matrix. $G_1, G_2$ and $G_3$ represent coherent displacements of the input vacua, $G_4$ and $G_6$ represent the cavity-qubit systems, and $G_5$ represents a beamsplitter modeling the lossy circulator. The output field $z(t)$ emerges from the output port of cavity 2 and is monitored by homodyne detection (see below). The SLH triples $(S, L, H)$ \cite{Gou.Jam-2007, Gou.Jam-2009, Gough:2012fl} for these blocks are
\bqa
G_1 &=& \left(1, \bar \epsilon(t), 0 \right) \nn \\
G_2 &=& \left(1, \bar A_d(t), 0 \right) \nn \\
G_3 &=& \left(1, \bar B_d(t), 0 \right) \nn \\
G_4 &=& \left(-\mathbf{1}_2, \left[\begin{array}{c} \sqrt{\kappa_1} \Op a \\ \sqrt{\gamma_1} \Op a \end{array}\right], \Delta_{1} \Op a\dg \Op a + \chi_1 \Op a\dg \Op a \Op \sigma_z^1 \right) \nn \\
G_5 &=& \left( \left[\begin{array}{cc} \sqrt{\eta_l} & i\sqrt{1-\eta_l} \\ i\sqrt{1-\eta_l}  & \sqrt{\eta_l}\end{array}\right], 0, 0 \right) \nn \\
G_6 &=& \left(-\mathbf{1}_2, \left[\begin{array}{c} \sqrt{\kappa_2} \Op b \\ \sqrt{\gamma_2} \Op b \end{array}\right], \Delta_{2} \Op b\dg \Op b + \chi_2 \Op b\dg \Op b \Op \sigma_z^2 \right) \nn
\eqa
where $\mathbf{1}_2$ is a $2\times 2$ identity matrix and $\eta_l$ is the 
efficiency of the circulator between the cavities (\ie $\eta_l=1$ implies no loss) \footnote{The zero (0) elements in SLH triples should be interpreted as zero matrices or vectors of the appropriate dimension.}. The SLH 
representation of the entire system is then formed by performing the following concatenation ($\boxplus$) and series ($\lhd$) products~\cite{Gou.Jam-2009}:
\bqa
(G_0 \boxplus G_6^{(1)} \boxplus G_0 \boxplus G_6^{(2)} ) \lhd  (G_5 \boxplus G_0 \boxplus G_0) \lhd \nn \\ (G_0 \boxplus G_4 \boxplus G_0) \lhd  (G_0 \boxplus G_1 \boxplus G_2 \boxplus G_3), 
\label{eq:slh_prod}
\eqa
where $G_0 = (1,0,0)$ is a pass-through component, and we have split the two ports of the second cavity as
\bqa
G_6^{(1)} &=& \left(-1, \sqrt{\kappa_2} \Op b, \Delta_{2} \Op b\dg \Op b + \chi_2 \Op b\dg \Op b \Op \sigma_z^2 \right) \nn \\
G_6^{(2)} &=& \left(-1, \sqrt{\gamma_2} \Op b, 0 \right) \nn
\eqa
for convenience (without this splitting, we would have to insert a routing element to swap the third and fourth signal lines after $G_4$ in Fig. \ref{fig:apparatus_slh}). The key assumption in this SLH representation of the entire network in terms of its components is that the fields propagate with negligible time delay between the components, which we assume to be true.

Evaluating the series and concatenation products in Eq.~(\ref{eq:slh_prod}) using the rules specified in Refs. \cite{Gou.Jam-2009, Gough:2012fl} yields the overall $G \equiv (S,L,H)$ for the network, from which an equation of motion for the two qubits and inter-cavity modes may be extracted (see Appendix). Adding phenomenological Markovian dephasing terms for the qubits with rate parameters $\gamma_d^i, i=1,2$, then results in the following master equation for the cavity mode and qubit degrees of freedom:
\bqa
\frac{\text{d}\varrho}{\text{d}t} &=& -i[H', \varrho] + \mathcal{L}_c \varrho+ \mathcal{L}_q \varrho \nn \\
\mathcal{L}_c \varrho&=&\mathcal{D}[\sqrt{\kappa_1(1-\etal)} \Op a]   + \gamma_1 \mathcal{D}[\Op a]\varrho  + \gamma_2 \mathcal{D}[\Op b]\varrho \nn \\
&&+ \mathcal{D}[-\sqrt{\kappa_1\etal}\Op  a  + \sqrt{\kappa_2}\Op b ]\varrho \nn\\
\mathcal{L}_q\varrho &=&   \sum_{i=1}^2 \gamma_d^i \mathcal{D}[\Op \sigma_z^i]\varrho,
\label{eq:full_me}
\eqa
where $\varrho$ is the combined density matrix of the two cavity modes and qubits, and $\mathcal{D}[A]B \equiv ABA\dg - \frac{1}{2} A\dg A B - \frac{1}{2} B A\dg A$. The effective Hamiltonian for the coupled system is
\bqa
H' &=&H_A+H_B+H_c'+H_d' \nn \\
H_c'&=& -i \frac{\kappa_{12}}{2}(\Op a\dg \Op b - \Op b\dg \Op a), \nn\\
H_d'&=&i\left(A_d(t)\Op a\dg -A_d^*(t)\Op a+B_d(t)\Op b\dg-B_d^*(t)\Op b\right),
\label{eq:H_me}
\eqa
where $\kappa_{12}=\sqrt{\kappa_1 \kappa_2 \etal}$, and the effective cavity drives are
\bqa
A_d(t)&=& \sqrt{\gamma_1}\bar A_d(t)+\sqrt{\kappa_1}\bar\epsilon(t)\nn\\
B_d(t)&=&\sqrt{\gamma_2}\bar B_d(t)-\sqrt{\kappa_2\eta_l}\bar\epsilon(t).\label{eq:drives}
\eqa
$H_c'$ describes the effective direct coupling of the two cavity modes due to the probe field that interacts with both cavities. Similarly, the effective drives for the cavity modes in $H_d'$ are composed of the probe field $\bar\epsilon(t)$ and the drive fields entering the input ports. The change in the sign of $\bar{\epsilon}(t)$ in the two expressions for effective cavity drives reflects the phase shift that the probe field picks up as it reflects from cavity 1.  Note that the coupling in $H_c'$ is reduced by the factor $\sqrt{\eta_l}$, due to losses in the circulator. Although the Hamiltonian form of this cavity coupling looks reversible, the irreversibility enforced by the circulator is nevertheless captured in the network model when the dissipative dynamics modeled by $\mathcal{L}_c$ is included. We shall see the effects of this explicitly below (see also Ref. \cite{Car-1993}). 

$\mathcal{L}_q$ accounts for intrinsic decoherence for the qubits, which is assumed to be pure dephasing dynamics.  In most of this work we neglect the contribution of relaxation (and heating) which typically contributes on a timescale $T_1$ much longer then the timescales of interest for establishing entanglement by continuous joint measurement, \eg \cite{Roch:2014ey}. However, in section \ref{sec:CW} we discuss how this effect may be included in the reduced model for the coupled qubit dynamics in the laboratory frame that is derived in section \ref{sec:red_sme}.

Each dissipator term in $\mathcal{L}_c$ takes into account the effect of a field irreversibly coupling out of the combined system. In particular, $\mathcal{D}[\sqrt{\kappa_1(1-\etal)} \Op a]$ accounts for photons lost between the two cavities, $\mathcal{D}[\Op a]$ accounts for the field emitted from the input port of cavity 1, $\mathcal{D}[\Op b]$ accounts for the field emitted from the input port of cavity 2, and $\mathcal{D}[-\sqrt{\kappa_1\etal} \Op a  + \sqrt{\kappa_2}\Op b ]$ accounts for the probe field that is sequentially reflected off the output ports of cavities 1 and 2. This final output channel is the only one that is monitored, and this correlated dissipator encodes the fact that when coherent light escapes the system, it has interacted with both the first and second cavity. It is therefore impossible to distinguish which cavity a decayed photon has come from, and thus it must be described using a combined operator.  
To complete the full model in the presence of measurement, we must describe the evolution of the system conditioned on homodyne measurement of the output field from cavity 2,
\beq
z(t) = -\sqrt{\kappa_1\etal}\Op  a(t)  + \sqrt{\kappa_2}\Op b(t).
\label{eq:zt}
\eeq
The homodyne measurement is implemented by mixing this signal field with a local oscillator of fixed phase reference, $\phi$, with respect to the initial phase of the probe field. This phase reference sets the measurement quadrature. The corresponding time evolution is given by \cite{Wiseman:2009vw}
\bqa
\frac{\text{d}\varrho}{\text{d}t} &=& -i[H', \varrho] + \mathcal{L}_c \varrho+ \mathcal{L}_q\varrho+ \mathcal{L}_m \varrho \nn \\
\mathcal{L}_m \varrho&=& \sqrt{\etam}\xi(t) \mathcal{H}[e^{i\phi}(-\sqrt{\kappa_1\etal} \Op a + \sqrt{\kappa_2}\Op b)]\varrho, 
\label{eq:sme}
\eqa
where $0<\etam \leq 1$ is the efficiency of the measurement, $\phi$ defines the measurement quadrature, and $\xi(t)$ is Gaussian white noise due to the measurement. This equation is in Ito form \cite{Gar.Zol-2004} and therefore $\xi(t)dt = dW(t)$, where $dW(t)$ is a Wiener increment satisfying $E\{dW(t)\}=0$ and $E\{dW(t)dW(s)\}=\delta(t-s)$ ($E$ denotes expectation value). The nonlinear conditioning superoperator $\mathcal{H}$ is defined as: $\mathcal{H}[A]B \equiv AB + BA\dg -\mathrm{Tr}(AB+BA\dg)B$. \erf{eq:sme} describes the conditioned state of the system under a homodyne measurement trace of the voltage
\beq
V(t) = \sqrt{\etam} \mathrm{Re} ( e^{i\phi}\langle-\sqrt{\kappa_1\etal} \Op a  + \sqrt{\kappa_2}\Op b \rangle )+ \xi(t),
\label{eq:voltage}
\eeq
where $\langle \Op A \rangle \equiv \tr(\Op A \varrho)$.
\erf{eq:voltage} expresses the monitored voltage in terms of the measured observable $z(t)$, which is a linear combination of intra-cavity field operators $\Op a$ and $\Op b$ (\erf{eq:zt}). 

The Heisenberg equations of motion for the expected values of the intra-cavity fields under the unconditioned evolution described by \erfs{eq:full_me}{eq:H_me} are:
\bqa
\dot{\expect{\Op a}} &=& -i \Delta_{1}  \expect{\Op a} - i\chi_1 \expect{\Op \sigma_z^1 \Op a} - \frac{\kappa_1 + \gamma_1}{2} \expect{\Op a} + A_d(t)  \nn \\
\dot{\expect{\Op b}} &=& -i\Delta_{2}  \expect{\Op b} - i\chi_2 \expect{\Op \sigma_z^2 \Op b} + \kappa_{12} \expect{\Op a} - \frac{\kappa_2 + \gamma_2}{2} \expect{\Op b} \nn \\
&& ~~~ + B_d(t). \nn
\label{eq:fieldeqs}
\eqa
These evolution equations make explicit the fact that the second cavity $\langle \Op b\rangle$ is driven by the first, $\langle \Op a\rangle$, but not vice-versa (\ie the irreversibility of the coupling between cavities). 
We assume that the driving fields ($\bar \epsilon(t), \bar{A}_d(t), \bar{B}_d(t)$) are all coherent states and therefore these expectation values are simply the coherent state amplitudes of the intra-cavity fields. We can write these coherent state amplitudes conditioned on the qubits being in specific states as:
\bqa
\dot{A}}^{(r) &=& -i\Delta_{1}  A^{(r)} - (-1)^r i \chi_1 A^{(r)} - \frac{\kappa_1+\gamma_1}{2} A^{(r)} \nn \\
&& ~~~+  A_d(t)  \nn \\
\dot{B}^{(rs)} &=& -i\Delta_{2}  B^{(rs)} - (-1)^s i\chi_2 B^{(rs)} - \frac{\kappa_2 + \gamma_2}{2} B^{(rs)} \nn \\
&& + B_d(t) + \kappa_{12} ~A^{(r)}.
\label{eq:cav_eqns}
\eqa
Here $A = \expect{\Op a}, B = \expect{\Op b}$ and the superscripts $r,s \in \{0,1\}$ indicate the conditioning on the state of the first and second qubit, respectively. The state of the second cavity, ${B}^{(rs)}$, is conditioned on the states of both qubits but the state of the first cavity, $A^{(r)}$, is only conditioned on the state of the first qubit, since there is no information flowing back from the second to the the first cavity. Explicitly, $A^{(11)}=A^{(10)}\equiv A^{(1)}$ and $A^{(01)}=A^{(00)} \equiv A^{(0)}$. These conditioned equations for intra-cavity amplitudes are linear and can be solved exactly (first solving for $A$ and then for $B$) for any values of the driving fields. Their exact solutions will be used below.

\subsection{Dynamics in the polaron frame}
\label{sec:polaron}
While the conditioned dynamical equation in \erf{eq:sme} is a full model of the experimental setup in Ref.~\cite{Roch:2014ey},  it is difficult to simulate since it involves both qubit and cavity degrees of freedom. Therefore it is convenient to derive an effective SME for the qubit degrees of freedom only. For this purpose, in this section, we develop an SME in a polaron frame where the average state of both intra-cavity fields is displaced to the vacuum. This SME is exact within the rotating wave approximation (RWA) and dispersive approximation implicit in \erf{eq:dispersive}, and becomes easy to simulate since the intra-cavity fields are always in the vacuum state. 

The polaron transformation provides a representation in which the cavity and qubit degrees of freedom are hybridized. The correct transform in this two cavity case is $\varrho^P(t) = U(t)\dg \varrho(t) U(t)$, with
\bqa
U(t) &= \sum_{i,j}\Op \Pi_{ij}D_1\left[A^{(i)}(t)\right]D_2\left[B^{(ij)}(t)\right],
\label{eq:polaron}
\eqa
where $\Op \Pi_{ij} = \ket{i}_1\bra{i}\otimes \ket{j}_2\bra{j}$ are projectors onto qubit states and $D_{1(2)}[X]$ is a displacement operator for cavity field $1(2)$, \ie
\bqa
D_1[X] &=& e^{X\Op a\dg - X^*\Op a} \nn \\
D_2[X] &=& e^{X\Op b\dg - X^*\Op b}. \nn
\eqa
For convenience, we define the following time-dependent qubit operators that depend on the intra-cavity field states:
\bqa
\Op \Pi_a(t) &\equiv & \Op \Pi_0 A^{(0)}(t) + \Op \Pi_1 A^{(1)}(t), \nn \\
\Op \Pi_b(t) &\equiv & \sum_{i,j={0,1}}\Op \Pi_{ij} B^{(ij)}(t), \label{eq:opfields}
\eqa
where $\Op \Pi_0 = \Op \Pi_{00}+\Op \Pi_{01}$ and $\Op \Pi_1 = \Op \Pi_{10} + \Op \Pi_{11}$. The quantities $A^{(r)}(t), B^{(rs)}(t), \Op \Pi_{a/b}(t)$ are all time-dependent, however, in the following we will often suppress the time parameter for the sake of brevity.
\erf{eq:opfields} can also be viewed as qubit projectors whose relative amplitudes depend on the cavity fields, which arises as a consequence of the hybridization of cavity and qubit degrees of freedom induced by the polaron transform. Note that with this definition, the polaron transformation can also be written as 
\beq
U(t) = D_1[\Op \Pi_a(t)] D_2[\Op \Pi_b(t)]
\eeq

The temporal evolution of the joint density matrix in this polaron frame is given by
\bqa
\frac{\text{d}\varrho^P}{\text{d}t} &=& \frac{\text{d}}{\text{d}t}U(t)\dg\varrho(t) U(t) \nn \\
&=& -i[H'^P, \varrho^P] + \mathcal{L}^P_c \varrho^P+ \mathcal{L}^P_q\varrho^P+ \mathcal{L}^P_m \varrho^P  \nn \\
&& - U\dg \dot{U} \varrho^P - \varrho^P\dot{U}\dg U, \nn
\eqa
where $H'^P, \mathcal{L}^P_c, \mathcal{L}^P_q, \mathcal{L}^P_m$ are as defined in \erf{eq:full_me} and \erf{eq:sme}, but with each operator transformed into the polaron frame according to $O^P = U(t)\dg O U(t)$. For example, transforming the field annihilation operators yields 
\bqa
U(t)\dg \Op aU  (t) &=& \Op a + \Op \Pi_a(t)\nn \\
U(t)\dg \Op bU(t) &=& \Op b+ \Op \Pi_b(t).
\eqa
Performing all the polaron transformations, the SME describing conditioned evolution in this frame then takes the explicit form (for an unnormalized density matrix):
\bqa
\frac{\text{d}\varrho^P}{\text{d}t} &=& -i[H_A + H_B +H_c' + H_q , \varrho^P] \nn\\
&&+ \mathcal{L}_c \varrho^P+ \mathcal{L}_q \varrho^P+ \mathcal{L}_q' \varrho^P+ \mathcal{L}_m \varrho^P  + \mathcal{L}_{mq} \varrho^P  \nn\\
&&+  \Op a[\varrho^P, \Gamma_1\Op \Pi_a\dg+\kappa_{12}\Op \Pi_b\dg] + [\Gamma_1\Op \Pi_a+\kappa_{12}\Op \Pi_b, \varrho^P] \Op a\dg   \nn\\
&&+  \Op b[\varrho^P, \Gamma_2 \Op \Pi_b\dg+ \kappa_{12}\Op \Pi_a\dg]+ [\Gamma_2\Op  \Pi_b+ \kappa_{12}\Op \Pi_a, \varrho^P] \Op b\dg, \nn\\
\label{eq:sme_polaron}
\eqa
where
\bqa
H_q&=& i\left(A_d(t)\Op \Pi_a\dg - A_d^*(t)\Op \Pi_a + B_d(t)\Op \Pi_b\dg  -B_d^*(t)\Op \Pi_b \right)\nn\\
\mathcal{L}_{q}'&=&\gamma_2 \mathcal{D}[\Op \Pi_b]+ (\kappa_1(1-\etal)+\gamma_1)\mathcal{D}[ \Op \Pi_a] \nn \\
&& +\mathcal{D}[-\sqrt{\kappa_1\etal} \Op \Pi_a + \sqrt{\kappa_2}\Op \Pi_b ]  \nn\\
\mathcal{L}_{mq}&=& \sqrt{\etam} \xi(t) \bar{\mathcal{H}}[e^{i\phi}(-\sqrt{\kappa_1\etal} \Op \Pi_a + \sqrt{\kappa_2}\Op  \Pi_b)], \label{eq:polaronterms}
\eqa
with $\Gamma_{i} = \gamma_{i}+\kappa_{i}$ being the total decay rate of cavity $i$, and $\bar{\mathcal{H}}[A]B \equiv AB+BA\dg$.
Comparison with \erf{eq:sme} shows that the polaron transformation has resulted in the additional terms $H_q, \mathcal{L}_{q}'$ and $\mathcal{L}_{mq}$ in the master equation. We sacrifice normalization in the following for simplicity, noting that the normalizing factor can always be recovered by computing $\tr(\frac{\text{d}\varrho^P}{\text{d}t})$.
In this polaron frame, there is no drive of the cavity modes by $A_d$ and $B_d$, because we are dynamically shifting the cavity states back to the vacuum. As a result, if the cavities are unpopulated initially, they remain unpopulated at all times. One way to see this is to note that all field operators act as annihilation operators on $\varrho^P$ in \erf{eq:sme_polaron}, and therefore there is no change in the states of the two cavity modes if they start in the vacuum. As a consequence, all terms involving cavity mode operators $\Op a$, $\Op b$ in \erf{eq:sme_polaron} have no effect and can be dropped.  This results in the following equation of motion for the qubit degrees of freedom in the polaron frame:
\bqa
\frac{\text{d}\varrho^P}{\text{d}t} &=& -i[H_q , \varrho^P] + \mathcal{L}_q \varrho^P + \mathcal{L}_q' \varrho^P+ \mathcal{L}_{mq} \varrho^P 
\label{eq:sme_polaron_red}
\eqa
The terms in $H_q$ represent Stark shifting of the energy levels due to the interaction with the measurement pulse.  The terms in $\mathcal{L}_{q}'$ represent information about the qubits leaking out (and thereby also dephasing the qubits) as a result of light exiting the various output ports in the system. $\mathcal{L}_{mq}$ represents stochastic measurement noise on the system as a result of the monitoring of one of the output ports.

It is important to note that \erf{eq:sme_polaron_red} contains no additional approximations beyond the RWA and the approximation of the Jaynes-Cummings interaction by the dispersive interaction between qubits and cavity modes. However, due to the polaron frame transformation, as long as the cavities are initially unpopulated, this equation efficiently simulates the coupled qubit and cavity degrees of freedom without the cost of keeping track of the quantized field states in the cavity. Instead, their influence is captured by the time-dependent operators $\Op \Pi_a$ and $\Op \Pi_b$ in \erf{eq:sme_polaron_red}.

\subsection{Transforming back to the lab frame}
\label{sec:reduced_lab}
In order to make predictions with respect to the lab frame, we transform the density matrix that results from \erf{eq:sme_polaron_red} back into the lab frame. We can achieve this by first noting that the state of the system at an arbitrary time in the polaron frame takes the form:
\beq
\varrho^P(t) = \sum_{ijkl} r_{ijkl}(t) \ket{ij}\bra{kl} \otimes \ket{00}\bra{00},
\label{eq:varrho_polaron_simple}
\eeq
where $r_{ijkl}(t) $ is  the solution to \erf{eq:sme_polaron_red}.
The $ijkl$ indices run over $\{0,1\}$ and index the qubit states. The second term in the tensor product, $\ket{00}\bra{00}$, is the state of the intra-cavity fields. Both modes are in the vacuum state since in the polaron frame the cavities remain unoccupied. Then the state of the entire system in the lab frame is given by $\varrho(t) = U(t) \varrho^P(t) U\dg(t)$, and the state of the qubits in the lab frame is given by
\beq
\rho(t) = \tr_{c1,c2}\left( U(t) \varrho^P(t) U\dg (t) \right),
\label{eq:red_tr}
\eeq
where the trace is taken over both cavity modes. Writing
\beq
\rho(t) = \sum_{ijkl} \rho_{ijkl}(t) \ket{ij}\bra{kl},
\label{eq:rho_qubits}
\eeq
we can determine the relation between $\rho_{ijkl}(t)$ and $r_{ijkl}(t)$, using the definition of the polaron transform in \erf{eq:polaron} and evaluating \erf{eq:red_tr}. We find that the diagonal elements remain unchanged by the transformation,
\beq
\label{eq:polaron_lab_pops}
\rho_{ijij}(t) = r_{ijij}(t), \nn
\eeq 
but that the off-diagonal components are modified as
\beq
\rho_{ijkl}(t) = r_{ijkl}(t) e^{\Upsilon_{ijkl}(t)},
\label{eq:comp}
\eeq
where the compensation factor is
\bqa
\Upsilon_{ijkl}(t) &=& i \mathrm{Im}\{A^{(k)*}A^{(i)}\} + i\mathrm{Im}\{B^{(kl)*}B^{(ij)}\} \nn \\
&&- \frac{|A^{(i)}-A^{(k)}|^2}{2} - \frac{|B^{(ij)}-B^{(kl)}|^2}{2}. 
\label{eq:compensation}
\eqa

These relations suggest that an efficient method for simulation of the system in the lab frame is to compute the time dynamics in the polaron frame according to \erf{eq:sme_polaron_red}, giving $r_{ijkl}(t)$, and then to compute the compensation to the off-diagonal elements given by \erf{eq:comp} at each time, to get the reduced state of the qubits in the lab frame. 

\subsection{Reduced equation of motion for the qubits}
\label{sec:red_sme}
Another approach for obtaining the state of the two qubits at any time is to formulate an equation of motion for just the qubit degrees of freedom in the lab frame. We begin with the expression for a general two qubit state in the lab frame given in \erf{eq:rho_qubits}. Taking the time derivative of this state yields:
\beq
\ddt \rho_{ijij}(t) = \ddt r_{ijij}(t) \nn
\eeq
for the diagonal elements, and 
\beq
\ddt \rho_{ijkl} = \left( \ddt r_{ijkl}(t) \right) e^{\Upsilon_{ijkl}(t)} + \rho_{ijkl}\left( \ddt \Upsilon_{ijkl}(t)\right) \label{eq:derivative_rho}
\eeq
The time derivative of $r_{ijkl}$ is determined by the polaron frame SME of \erf{eq:sme_polaron_red}, and the time derivative of the compensation factor can easily be found from its definition in \erf{eq:compensation} together with the equation of motion for the conditional intra-cavity fields, \erf{eq:cav_eqns}. This is similar to the approach taken in Ref. \cite{Gam.Bla.etal-2008} for a single cavity setup. 

Computing the derivatives required in \erf{eq:derivative_rho} and canceling common factors yields the following equations of motion for the components of the (unnormalized) lab frame qubit density matrix:
\bqa
\dot{\rho}_{ijkl} &=& \rho_{ijkl}\left[ i2\chi_1(1-\delta_{ik})\left((-1)^{i} A^{(k)*}A^{(i)} \right) \right. \nn \\
&&~~~~~~~ +i2\chi_2(1-\delta_{jl})\left((-1)^{j} B^{(kl)*}B^{(ij)} \right) \nn \\
&&~~~~~~~ -2\gamma_d^1(1-\delta_{ik}) - 2\gamma_d^2(1-\delta_{jl}) \nn \\
&&~~~~~~~\left. +\sqrt{\eta_{\rm m}} \xi(t) \{\tilde{B}^{(ij)} + \tilde{B}^{(kl)*} \}\right],
\eqa
with 
\beq
\tilde{B}^{(ij)} \equiv e^{i\phi}(-\sqrt{\kappa_1\eta_{l}}A^{(i)} + \sqrt{\kappa_2}B^{(ij)}),
\eeq
being the conditional output fields. This evolution can also be represented in matrix form (again for an unnormalized qubit density matrix) as:
\bqa
\frac{\text{d}\rho}{\text{d}t} &=& \sum_{ijkl} a_{ijkl}(t) \Op \Pi_{ij} \rho(t) \Op \Pi_{kl}
 + \mathcal{L}_q \rho(t) + \mathcal{L}_{mq} \rho(t) 
\label{eq:qubit_sme}
\eqa
with 
\bqa
a_{ijkl}(t) &\equiv& i2\chi_1(1-\delta_{ik})\left((-1)^{i} A^{(k)*}A^{(i)} \right) \nn \\
&&+i2\chi_2(1-\delta_{jl})\left((-1)^{j} B^{(kl)*}B^{(ij)} \right) \nn
\eqa
\erf{eq:qubit_sme} should be treated with care since although it looks like an SME with the deterministic component in Lindblad form, it is not strictly in Lindblad form because the coefficient matrix defined by $a_{ijkl}(t)$ is not necessarily positive. Hence this equation can result in non-Markovian evolution of the qubit. Physically, this is a result of tracing out the intracavity degrees of freedom that are strongly entangled with the qubit states. Therefore physical interpretation of the rate coefficients $a_{ijkl}(t)$ in \erf{eq:qubit_sme} is difficult. Nevertheless, \erf{eq:qubit_sme} does generates the correct qubit evolution in the lab frame and presents an alternative to simulating the qubit evolution in the polaron frame according to \erf{eq:sme_polaron_red}.

\subsection{Simulation in different frames}
\label{subsec:trajectories}

\begin{figure}
\centering \includegraphics[width=1.0\columnwidth]{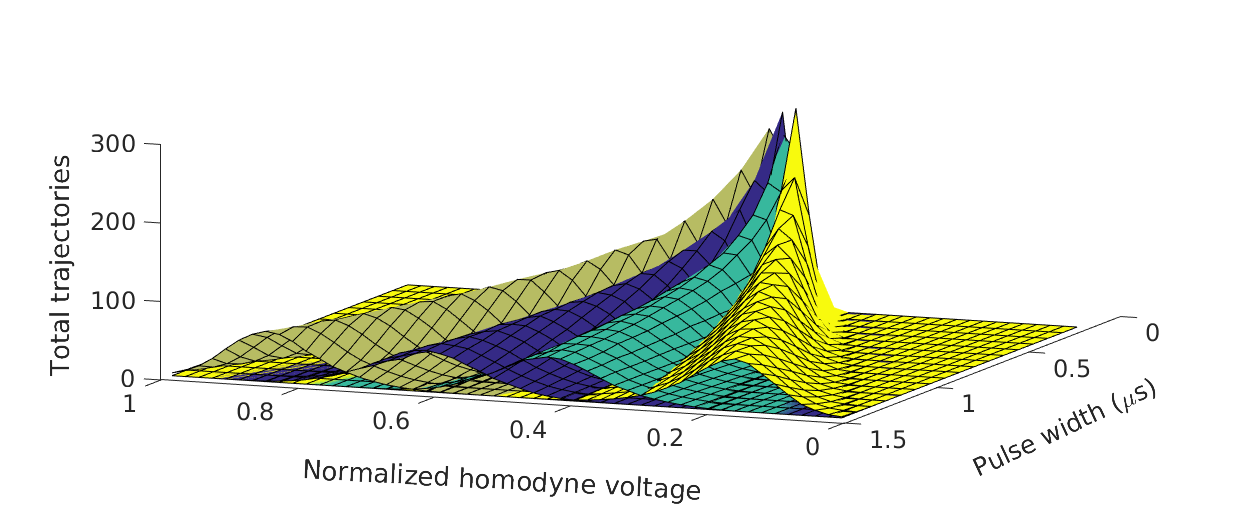}
\caption{(Color online) Distribution of qubit state populations (200,000 trajectory simulations) as a function of normalized homodyne voltage and measurement pulse width,
for simulation parameters corresponding to slightly different qubits. See section \ref{subsec:trajectories} for the parameter values. The initial state is given in \erf{eq:init}. The z-axis is the total number of trajectories that result in the state $|00\rangle$ (brown), $|01\rangle$(blue), $|10\rangle$(green), or  $|11\rangle$ (yellow).  \label{fig:PvsTvsJnoparity}}
\end{figure}
\begin{figure}
\centering \includegraphics[width=1\columnwidth]{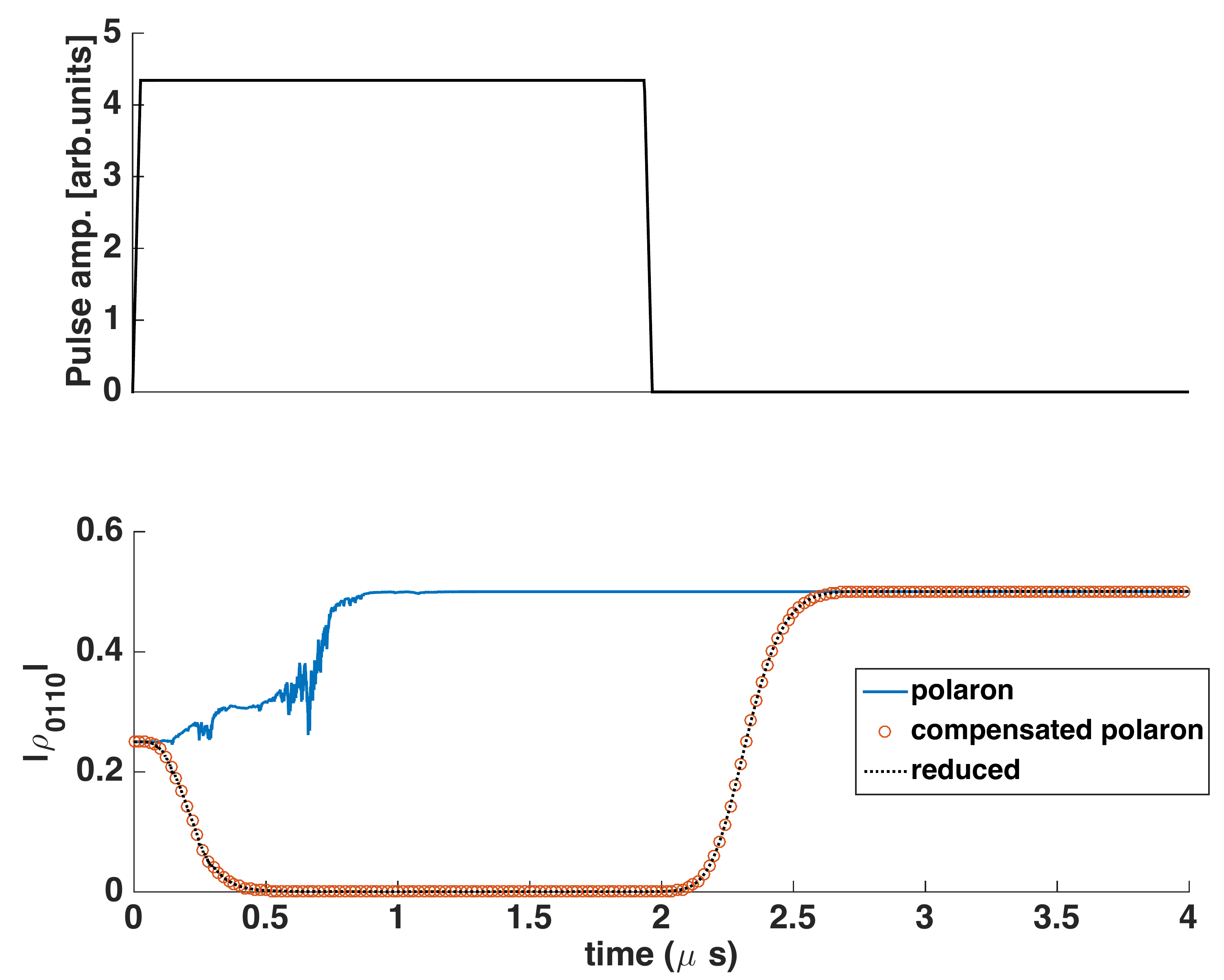}
\caption{(Color online) Bottom panel: Sample trajectory of $|\rho_{0110}(t)|$, the absolute value of an off-diagonal qubit density matrix element simulated using the three dynamical equations derived in section \ref{sec:polaron} (polaron, \erf{eq:sme_polaron_red}), \ref{sec:reduced_lab} 
(lab frame compensated polaron, Eqns.~(\ref{eq:red_tr})-(\ref{eq:compensation}), and \ref{sec:red_sme} (reduced equation of motion, \erf{eq:qubit_sme}). We use the ideal setting with identical cavities and lossless transmission.  All parameters are described in the main text. Top panel: measurement pulse $A_d(t)$ for this trajectory.} \label{fig:mes_traj}
\end{figure} 

Simulation of the system dynamics using any one of these three dynamical equations allows the generation of statistics for the homodyne voltage measured in individual runs of an experiment.  As discussed above, in both the polaron and laboratory frames (Eqs.~(\ref{eq:sme_polaron_red}) and  (\ref{eq:qubit_sme})), the measurement observable corresponds to 
\beq
\mathrm{Re} \{e^{i\phi}\expect{-\sqrt{\kappa_1\etal} \Op \Pi_a + \sqrt{\kappa_2}\Op  \Pi_b}\}.\nn
\eeq

Fig.~\ref{fig:PvsTvsJnoparity} demonstrates the distribution of measurement outcomes as a function of the normalized homodyne voltage and the measurement pulse width. We used the polaron frame reduced master equation, \erf{eq:sme_polaron_red}, with a large number of random realizations of $dW(t)$ (200,000 trajectories) to obtain these results. This choice of frame is made here for computational efficiency. For these simulations we use the parameters $\frac{\chi_1}{2\pi}=2$MHz, $\frac{\chi_2}{2\pi}=1.0$MHz,  $\frac{\kappa_1}{2\pi}=18$MHz, $\frac{\kappa_2}{2\pi}=22$MHz, $\Delta_1=\Delta_2=0$, $\gamma_1= \frac{\kappa_1}{20}$,$\gamma_2= \frac{\kappa_2}{20} $, $\sqrt{\gamma_1}A_d(t)=13$MHz, $\eta_l=0.9$, and $\eta_{\rm m}=0.4$. Given an initial state that is the equal superposition of the computational basis states, 
\beq
\ket{\Psi_0} = \frac{1}{2}(\ket{0} + \ket{1})\otimes (\ket{0} + \ket{1}), \label{eq:init}
\eeq
continuous measurement will eventually collapse onto one of the basis states and the corresponding measurement voltage will be distributed according to one of the four Gaussian distributions in Fig. \ref{fig:PvsTvsJnoparity}. The smaller the relative phase shift of the coherent states entangled to different qubits levels, the longer it will take for the Gaussian distributions to separate and to clearly distinguish the basis states.

In contrast to the diagonal elements of the density matrix, which as stated above are independent of the frame in which the simulation is made, the coherences of the qubit populations will differ greatly between frames. Here we present an example time evolution that illustrates the points raised in the derivation of equations of motion for the qubits in the lab frame
in sections~\ref{sec:reduced_lab} and \ref{sec:red_sme}.
We assume no loss and identical cavities for simplicity here (we shall refer to this as the ``ideal setting"), with simulation parameters: $\eta_l=1, \frac{\kappa_1}{2\pi}=\frac{\kappa_2}{2\pi}=1.5 \rm{MHz}, \frac{\chi_1}{2\pi}=\frac{\chi_2}{2\pi}=0.5 \rm{MHz}, \Delta_1=\Delta_2=0$ and $\gamma_1=\gamma_2=0$. Fig. \ref{fig:mes_traj} shows the $\rho_{0110}$ off-diagonal component of the qubit density matrix under a particular measurement trajectory, simulated using the three dynamical equations derived above: \emph{polaron} \erf{eq:sme_polaron_red}, \emph{compensated polaron} Eqns.~(\ref{eq:red_tr})-(\ref{eq:compensation}), and \emph{reduced} \erf{eq:qubit_sme}. The top panel of the figure also shows the measurement pulse $A_d(t)$ that produced the trajectory. We choose $\bar A_d(t) = \bar B_d(t) = 0$ since it is the appropriate drive of the two cavities in this ideal parameter regime (see section \ref{sec:ent_condition_id}). As this figure shows, and as expected from the derivation in section \ref{sec:red_sme}, the compensated polaron and the reduced evolution equations produce exactly the same results. However, both of these differ from the results of the polaron frame evolution, \erf{eq:sme_polaron_red}, unless the photon population of both cavities is zero, which occurs only at $t=0$ and at long times after the measurement pulse has ended. 
Another interesting feature is the revival of coherence produced by the two lab frame SMEs (polaron compensated and reduced) between $2$ and $2.5 \mu$s, which is after the pulse has decayed to zero. This is a signature of the non-Markovian nature of the evolution equations resulting from elimination of the strongly entangled cavity states.  Note that a similar effect should be expected even in the single cavity case \cite{Gam.Bla.etal-2008}, and even in that simpler case, the stated lab ``dephasing rate" would be expected to be non-Markovian, and lead to non-decaying coherence in some parameter regimes. Therefore, care should be taken with the interpretation of parameters entering laboratory frame qubit SMEs derived in this fashion.

\section{Conditions for generating entanglement}
\label{sec:entanglementcondition}
Having developed a model for the apparatus in Fig. \ref{fig:apparatus}, we now demonstrate how a sequential coherent probe and subsequent homodyne detection can result in a half-parity measurement of the qubits, which in turn can probabilistically entangle the qubits. The derivation of a reduced equation of motion under continuous weak measurement for the qubit degrees of freedom alone, \erf{eq:qubit_sme}, allows us to identify the exact qubit observable that is being monitored by the sequential probe, namely $\mathrm{Re} \{e^{i\phi}\expect{-\sqrt{\kappa_1\etal} \Op \Pi_a + \sqrt{\kappa_2}\Op  \Pi_b}\}$. In this section, we will specify how the parameters that can be controlled \emph{in-situ} (\eg frequency, amplitude and phase of drive tones $ A_d(t),  B_d(t)$) can be tuned such that monitoring this observable can generate entanglement between the qubits even when the system parameters are not ideal.

The key fact that the probabilistic entanglement scheme relies on is that under the half-parity measurement, the states $\ket{01}$ and $\ket{10}$ are indistinguishable. In that situation, starting in the initial separable state in \erf{eq:init}, for entanglement to be generated by measurement the initial coherence between $\ket{01}$ and $\ket{10}$ must be preserved (or at least not decay substantially). For this to happen, the indistinguishability between these states must be maintained at all times, \ie it is not sufficient that the measurement voltage be the same for both states merely at the final time. Specifically, we require that the monitored observable have the same value when the qubits are in state $\ket{01}$ or in state $\ket{10}$, or equivalently, for all $t$, $\mathrm{Re} \tilde{B}^{(01)}(t)=\mathrm{Re} \tilde{B}^{(10)}(t)$, which is guaranteed if
\bqa
&&\tr\big[(-\sqrt{\kappa_1\etal} \Op \Pi_a(t) + \sqrt{\kappa_2}\Op \Pi_b(t))\ket{01}\bra{01}\big] \nn \\
&& = \tr\big[(-\sqrt{\kappa_1\etal} \Op \Pi_a(t) + \sqrt{\kappa_2}\Op \Pi_b(t))\ket{10}\bra{10}\big] \nn \\
\Rightarrow && \tr\big[ (-\sqrt{\kappa_1\etal} \Op \Pi_a(t) + \sqrt{\kappa_2}\Op \Pi_b(t)) \Op \Xi \big] =0,
\label{eq:halfparity}
\eqa
where $ \Op \Xi \equiv \ket{01}\bra{01} - \ket{10}\bra{10}$. 
We note that derivatives with respect to time of the expression on the left should ideally also be zero (since we demanding that the condition holds for all $t$). We will use the derivatives of ${\Op \Pi}_a$ and $\Op \Pi_b$ in the following and so we explicitly write the first derivatives of these operators here:
\bqa
\dot{\Op \Pi}_a(t) &=& -\tilde{\kappa}_1 \Op \Pi_a(t) - i\chi_1 \Op\sigma_z^1 \Op \Pi_a(t)+ A_d(t) \nn\\
\dot{\Op \Pi}_b(t) &=& -\tilde{\kappa}_2 \Op \Pi_b(t) + \kappa_{12}\Op \Pi_a(t) -i\chi_2 \Op \sigma_z^2 \Op \Pi_b(t) \nn \\
&& ~~~~~~ + B_d(t)
\label{eq:pib_dot}
\eqa
with $\tilde\kappa_i=\kappa_i/2+\gamma_i/2+i\Delta_i$.  These equations were obtained by using the definitions of the operators in \erf{eq:opfields} and the conditional cavity state equations of motion in \erf{eq:cav_eqns}. The second derivatives of $\Op \Pi_a$ and $\Op \Pi_b$ can be obtained in the same manner.

To derive a prescription for tuning the experimental parameters (specifically the compensating field, $B_d(t)$), we write $\Op \Pi_b$ in terms of its first derivative using \erf{eq:pib_dot} and substitute the result into \erf{eq:halfparity}, to obtain
\bqa
&&\tr\left[ \left( ( \kappa_{12}\Op\Pi_a+B_d-\Op{\dot\Pi_b})\frac{\tilde\kappa_2-i\chi_2\Op\sigma_z^2}{\tilde{\kappa_2}^2+{\chi_2}^2} \right. \right. \nn \\
&& ~~~~~~~~~~ \left. \left. -\frac{\sqrt{\kappa_1\etal}}{\sqrt{\kappa_2}}\Op\Pi_a \right) \Op\Xi\right]=0,
\label{eq:halfparityexpl}
\eqa
This is a general dynamical condition that the parameters in the system need to satisfy as the system evolves. However, this is a self-consistency equation for $B_d$ because the operator $\dot{\Op \Pi_b}$ depends implicitly on $B_d$, and thus it does not provide an explicit solution for the compensating field.  In the following, we discuss simple explicit solutions of \erf{eq:halfparityexpl} for three limiting, but physically relevant, regimes, as well as the more complex general solution that requires shaped pulses.

\subsection{Adiabatic regime}
\label{sec:ent_condition_ad}
Consider the regime where the probe fields vary very slowly or not at all (\eg steady-state or continuous-wave measurement), specifically, this is the limit where  $\dot A_d,\dot B_d\ll \kappa_1,\kappa_2$. In this case, we can approximate $A_d(t)$ and $B_d(t)$ as constant fields for short times and solve for the ``adiabatic values" of the hybridized field operators by setting the derivatives in equations \erf{eq:cav_eqns} to zero, resulting in 
\bqa
\Op\Pi_a^{ad}(t)&=&\frac{A_d(t)(\tilde\kappa_1-i\chi_1\Op\sigma_z^1)}{\tilde{\kappa_1}^2+{\chi_1}^2} \nn \\
\Op\Pi_b^{ad}(t)&=&\frac{(\kappa_{12}\Op\Pi_a^{ad}(t)+B_d(t))(\tilde\kappa_2-i\chi_2\Op\sigma_z^2)}{\tilde{\kappa_2}^2+{\chi_2}^2}.
\label{eq:steadystates_general}
\eqa
Substituting $\Op \Pi_a \rightarrow \Op \Pi_a^{ad}$ in \erf{eq:halfparityexpl}, dropping the $\Op{\dot\Pi_b}$ term (since this is small in this regime), and then solving for $B_d(t)$ in terms of the other quantities, yields
\beq
B_d^{ad}(t) =\kappa_{12}A_d(t)\frac{\left(\chi_1\tilde\kappa_2-\chi_2\tilde\kappa_1\right)-(\tilde{\kappa_2}^2+{\chi_2}^2)\chi_1/\kappa_2}{\chi_2(\tilde{{\kappa}_1}^2+{\chi_1}^2)}.
\label{eq:entangcond}
\eeq
This equation defines the value of $\bar{B}_d(t)$, the compensation field driving the second cavity, that achieves the desired indistinguishability of the states $\ket{01}$ and $\ket{10}$.  Equivalently, another approach to achieving indistinguishability, without using the compensating drive (i.e., setting $\bar{B}_d(t)=0)$, is to tune the frequency of the drive $A_d$. This gives a condition on $\Delta_1(t)$ in order to meet \erf{eq:entangcond}, and was the approach chosen in Ref.~\cite{Roch:2014ey} due to its simplicity. However we note that this  is only possible for certain parameter ranges. In both cases, \erf{eq:entangcond} can be met by tuning the parameter(s) \emph{in-situ} to obtain the same measurement statistics when the qubits are in state $\ket{01}$ and in state $\ket{10}$. 

\begin{figure}
\centering \includegraphics[width=0.9\columnwidth]{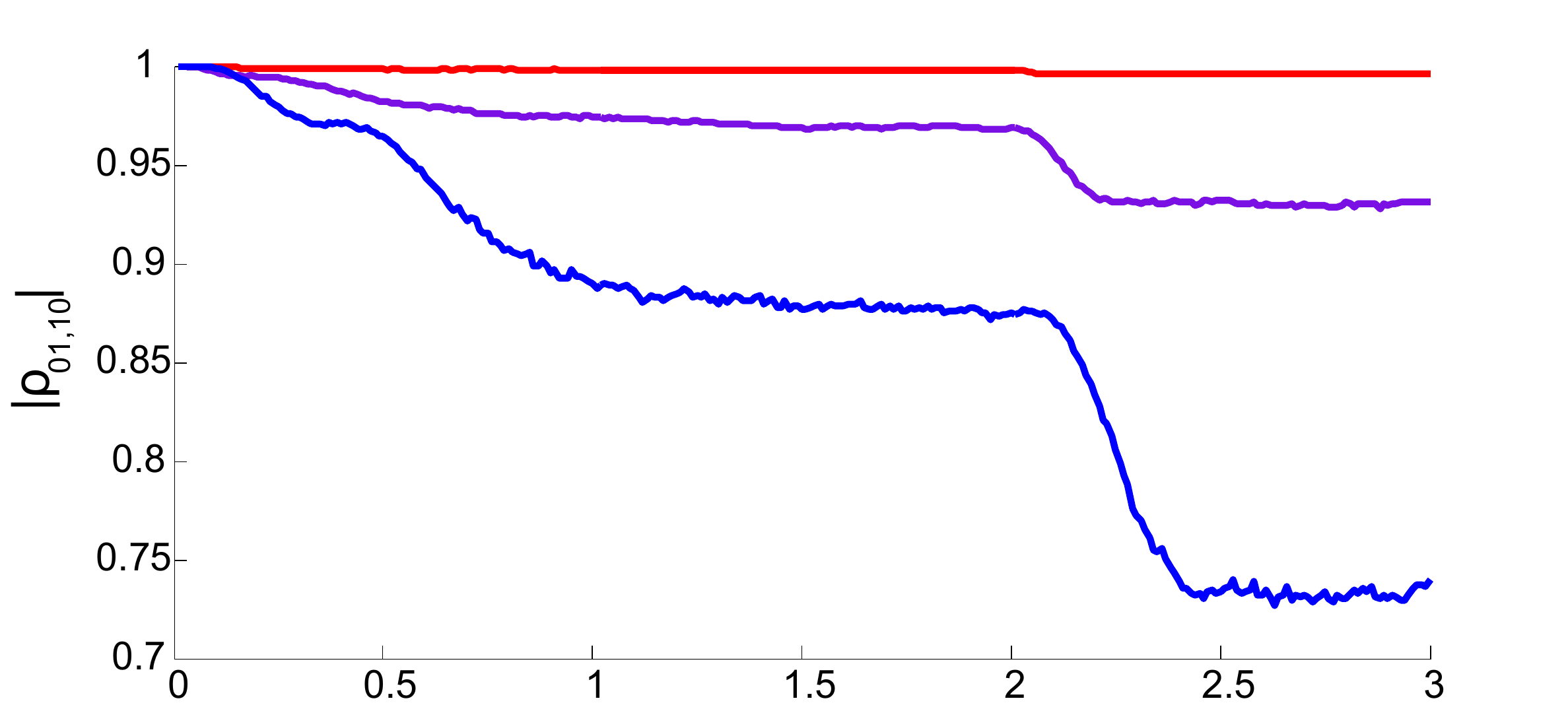}
\centering \includegraphics[width=0.9\columnwidth]{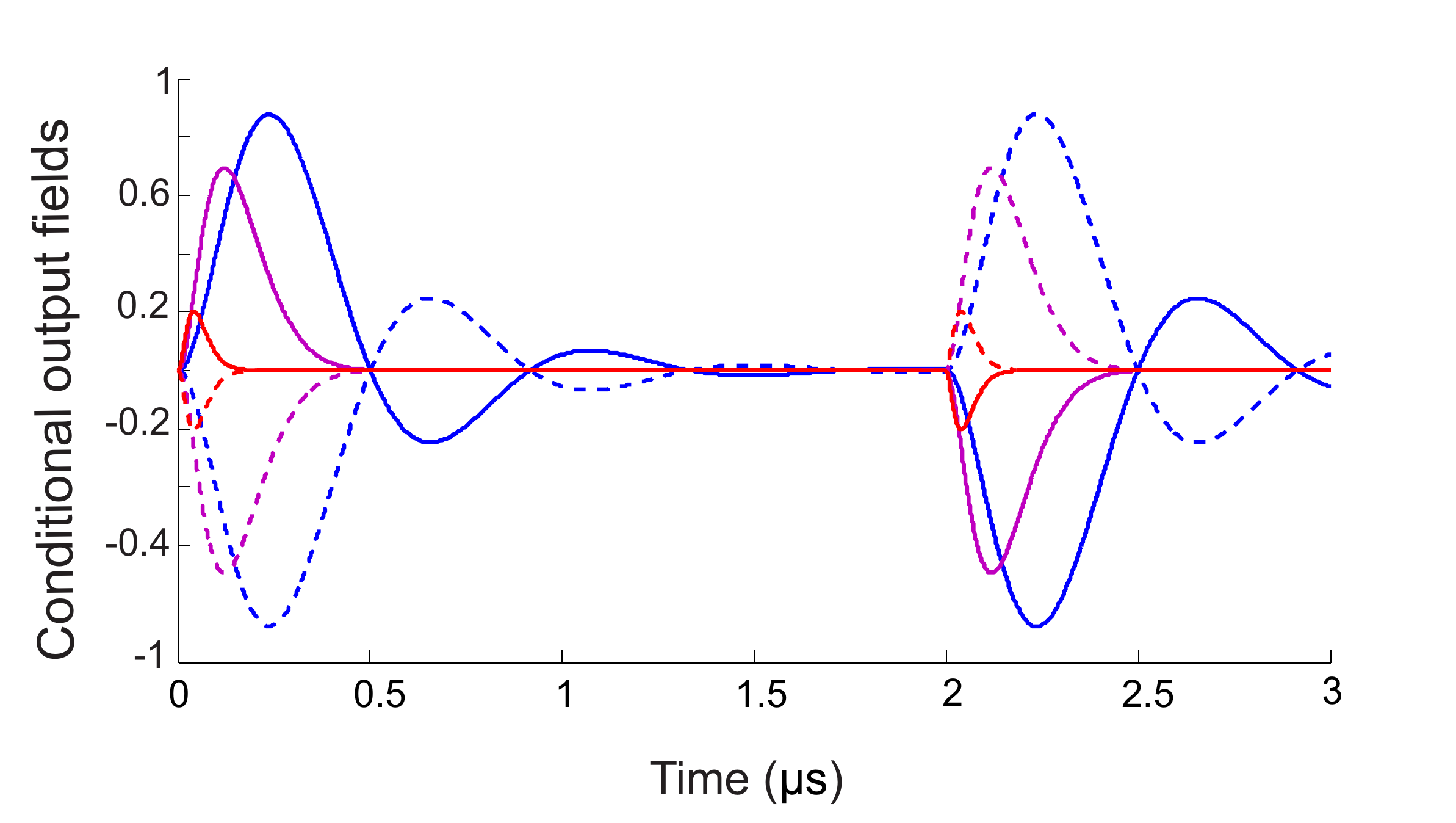}
\caption{(Color online) 
Top panel: Coherence between the $\ket{01}$ and $\ket{10}$ states. The blue (lower), purple (middle), and red (upper) lines correspond to $\kappa_1=\{1,5,17\}$MHz , respectively, with $\frac{\kappa_2}{2\pi}=\frac{\kappa_1}{2\pi}+2.5$ in all cases.  Bottom panel: Conditional output field amplitudes when the qubits are in states $\ket{01}$ ($\mathrm{Re}\{ \tilde B^{(01)}\}$, solid) and $\ket{10}$ ($\mathrm{Re}\{ \tilde B^{(10)}\}$, dashed), for different values of the cavity decay rate $\kappa_1$.   All other parameters are described in the main text.
\label{fig:transients}}
\end{figure}

\subsection{Bad cavity limit}
The adiabatic approach works best in the limit of very large cavity decay rates $\kappa_i$, where the transient-evolution time periods leading up to and following the steady-state are short enough enough to be entirely negligible.  In such a case, we can instead simply use a square measurement pulse $A_d$, as well as for the compensation via auxiliary measurement drive $B_d$ or the frequency calibration $\Delta_1$.

If we additionally assume  $\kappa_1,\kappa_2 \gg \chi_1,\chi_2$ (\ie take the bad cavity limit), \erf{eq:entangcond} further simplifies to 
\beq
B_d^{uni}(t) =-\sqrt{ \frac{\eta \kappa_1}{\kappa_2}}A_d(t)\frac{  (\tilde \kappa_2 ( \Delta_2(t)-\kappa_2/2) \chi_1+\tilde \kappa_1 \kappa_2 \chi_2/2)}{ \tilde{ {\kappa}_1}^2 \chi_2}.
\label{eq:entangcond2}
\eeq
Here we have retained the time dependence for generality but this is typically not necessary. The simplicity of this approach makes it of great practical utility and as such it was used in Ref.~\cite{Roch:2014ey}. Note however, that the cavity decay rates can only be increased up to a certain point imposed by physical constraints and so a small transient error may remain.

Fig.~\ref{fig:transients} (upper panel) shows the degradation of coherence as a result of such transient populations in the cavities. These transients exist because we have used \erf{eq:entangcond} (which reduces to \erf{eq:entangcond2} for larger values of cavity decay rates) for compensation. This figure shows the decay of the off-diagonal element $|\rho_{0110}|$ while the measurement pulse is applied, for $\frac{\kappa_1}{2\pi}$ taking on values $\{1,5,17\}$, with $\frac{\kappa_2}{2\pi}$=$\frac{\kappa_1}{2\pi}+2.5$. $\frac{\chi_1}{2\pi}=1.2$MHz and all other parameters are the same as in Fig.~\ref{fig:mes_traj}. The total measurement time required is kept approximately fixed by setting $A_d=\frac{\sqrt{0.9\kappa_1}}{2\pi}$ and employing the same pulse shape as in Fig.~\ref{fig:mes_traj}.  These calculations employed the polaron frame dynamical equation \erf{eq:sme_polaron_red}, with initial state $\ket{\Psi_0}$ and averaging over 1500 trajectories that resulted in the desired outcome. The bottom panel of the figure shows the conditional output fields $\mathrm{Re}\{ \tilde B^{(01)}\}$ and $\mathrm{Re}\{ \tilde B^{(10)}\}$ corresponding to the qubits being in state $\ket{01}$ and $\ket{10}$, respectively\footnote{We have chosen $\phi=\pi/2$, which is shown below to be the correct quadrature to measure for the half parity measurement.}.  Ideally, these conditional output fields should be identical at all times. However, as expected the qubit states are distinguishable by the output fields during the pulse transients, since the simplified compensation prescribed by \erf{eq:entangcond} (or \erf{eq:entangcond2} for the larger values of cavity decay $\kappa_{1,2}$) is utilized. Furthermore, the greater this distinguishability, the greater the associated loss of coherence. Fig.~\ref{fig:transients} shows that with larger cavity decay rates, these transients become smaller; a cavity decay rate of $\kappa_1=1$MHz causes $27\%$ loss of coherence, while in contrast a value $\kappa_1=17$MHz causes negligible loss.

\subsection{Ideal system parameter regime}
\label{sec:ent_condition_id}
In the ideal case, when the transmission is lossless ($\eta_l$=1) and the cavities and qubits are identical ($\kappa_1$=$\kappa_2$=$\kappa$, $\chi_1$=$\chi_2$=$\chi$, $\Delta_1$=$\Delta_2$=$0$,$\gamma_1$=$\gamma_2$=$\gamma$), \erf{eq:entangcond} reduces to 
\beq
B_d(t)=-A_d(t).\label{eq:entangcond3}
\eeq
This is equivalent to using the probe field together with only the reflection mode of the cavities, such that the compensation fields are not needed, i.e., $\bar A_d(t)$=$\bar B_d(t)$=$0$ in \erf{eq:drives}. We note that the same result is obtained if both cavities are driven only through their input ports, with $\sqrt{\gamma_2}\bar B_d(t)=-\sqrt{\gamma_1}\bar A_d(t)$, and $\bar\epsilon(t)$=$0$, i.e., the probe field is absent. 

In this ideal parameter case it is especially simple to see how the sequential probe reproduces the half-parity measurement. Explicitly, we observe that the adiabatic compensation in this ideal case results in \erf{eq:steadystates_general} reducing to 
\bqa
\Op\Pi_a^{ad,ideal}(t)&=&\frac{A_d(t)(\kappa/2-i\chi\Op\sigma_z^1)}{\kappa^2/4+\chi^2}\nn\\
\Op\Pi_b^{ad,ideal}(t)&=&\frac{(\kappa\Op\Pi_a^{ad,ideal}(t)-A_d(t))(\kappa/2-i\chi\Op\sigma_z^2)}{\kappa^2/4+\chi^2}\nn\\
\eqa
where we have set $\gamma_i=0$, since the fields entering the input ports are zero in this case.

The qubit observable being monitored in this ideal case is then
\bqa
&\mathrm{Re}& \expect{e^{i\phi}\sqrt{\kappa}(-\Op \Pi_a^{ad,ideal} + \Op \Pi_b^{ad,ideal})} \nn \\
&&=\frac{\sqrt{\kappa}}{d} \mathrm{Re} \Big\{ e^{i\phi}A_d(t)\left( \left[\frac{\kappa^3}{4d}-\frac{\kappa}{2}\right] - \frac{\kappa\chi^2}{d}\expect{\Op\sigma_z^1\Op\sigma_z^2} \right. \nn \\
&& \left.+ i\left[\chi-\frac{\kappa^2 \chi}{2d} \right]\expect{\Op\sigma_z^1+\Op\sigma_z^2}\right),
\label{eq:obs_ideal}
\eqa
where $d \equiv \kappa^2/4+\chi^2$. Choosing $A_d(t)$ real and setting the homodyne phase $\phi=\frac{\pi}{2}$ results in a measurement $\propto \expect{\Op\sigma_z^1 + \Op\sigma_z^2}$. 

We note that this ideal case affords the significant benefit that, the transients will exactly cancel for all parameter values and the indistinguishability condition (\erf{eq:halfparity}) will be met at all times. This is apparent in Fig.~\ref{fig:mes_traj}, where we see that in this setting, the coherence of the entangled state does not decrease at all during the periods of transient evolution of the cavities. The reason for this can easily be understood by looking at the equations of motion for the cavity dependent qubit projectors, \erf{eq:pib_dot}, in the limit of identical cavity parameters and using \erf{eq:entangcond3} for the ideal setting. Then, the condition for indistinguishability for identical cavities (i.e. for $\mathrm{Re}\{ e^{i\phi}(\Op\Pi_b(t) - \Op\Pi_a(t))\}$) differentiated twice
\bqa
&\mathrm{tr}& \left[\left(\Op{\ddot\Pi}_b(t)-\Op{\ddot\Pi}_a(t) \right) \Op\Xi\right]= 0 \nn 
\eqa
gives, after substituting the derivative of \erf{eq:pib_dot} followed by  \erf{eq:pib_dot} itself and \erf{eq:entangcond3}:
\bqa
&\mathrm{tr}& \left[\left( \chi^2(\Op{\dot\Pi}_b(t)-\Op{\dot\Pi}_a(t))-\frac{\kappa^2}{4}(\Op\Pi_b(t)-\Op\Pi_a(t)) \right) \Op\Xi\right] =0 \nn
\eqa
From this equation it is clear that if the condition is met initially, it is met also at all later times. This provides significant motivation to make the parameters for the two cavities as similar as possible and to make the transmission between cavities lossless, i.e., $\eta_l=1$).

\subsection{General dynamic condition}
\label{subsec:dyanmicentanglement}
\begin{figure}
\centering \includegraphics[width=0.99\columnwidth]{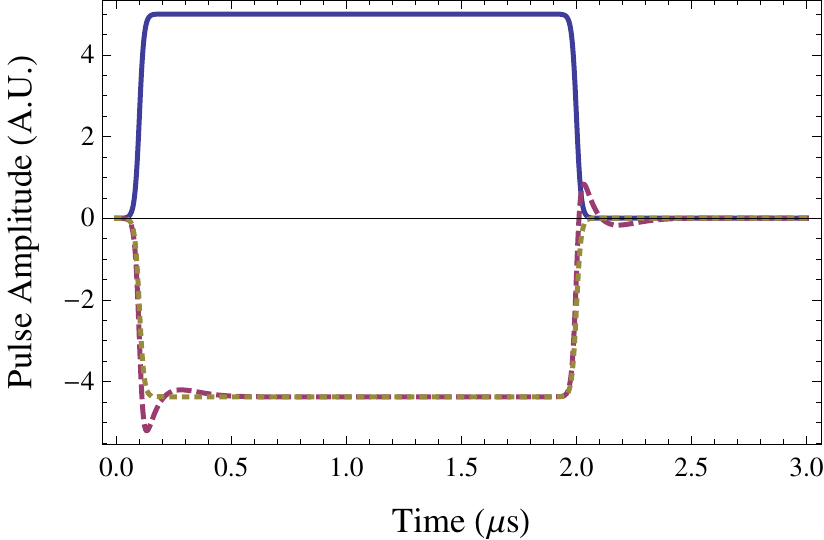}
\caption{(Color online) Example of a measurement pulse $A_d(t)$ and compensation pulses $B_d(t)$ for two cavities with identical frequencies but different transmittivities, $\frac{\kappa_1}{2\pi}$=3.9MHz and $\frac{\kappa_2}{2\pi}$=3.5MHz.  All other parameters are the same as in Fig. 3.  Blue solid line: pulse $A_d(t)$.  Orange dotted line: compensation pulse $B_d(t)$ when the adiabatic approximation is used (\erf{eq:entangcond}). Red dashed line: exact solution given by \erf{eq:dynamicBd}.  }
\label{fig:pulseshaping}
\end{figure}

Achieving the ideal setting of symmetric cavities and lossless transmission between them is experimentally challenging.  In this subsection we show how to eliminate the detrimental effects of transients even in nonideal cases by shaping the measurement and compensation pulses. In particular, we show how the compensation field $B_d(t)$ can be shaped to ensure that the indistinguishability condition, \erf{eq:halfparity}, is met at all times. 

In the following, we leave $A_d(t)$ unchanged (thus fully specifying $\Op\Pi_a(t)$) and shape the compensation field by adding a component $\Delta B_d(t)$ so that $B_d(t)=B_d^{ad}(t)+\Delta B_d(t)$ enforces the entanglement criteria, \erf{eq:halfparity}, at all times.  To derive the form of $\Delta B_d(t)$,
we supplement \erf{eq:halfparity} with its first and second derivative, which must also be equal to zero.

The second derivative of \erf{eq:halfparity} gives, upon inserting the derivative of \erf{eq:pib_dot}, 
\beq
\tr \left[ \left(\sqrt{\frac{\kappa_1\etal}{\kappa_2}} \Op{\ddot \Pi}_a- \kappa_{12}\Op{\dot\Pi}_a+(\tilde\kappa_2+i\chi_2\Op\sigma_z^2)\Op{\dot\Pi}_b \right) \Op\Xi\right]=0, \nn
\eeq
while the first derivative of \erf{eq:halfparity} gives simply
\beq
2\tilde\kappa_2\tr\left[ \left(\sqrt{\frac{\kappa_1\etal}{\kappa_2}}\Op{ \dot\Pi}_a-\Op{\dot\Pi}_b\right) \Op\Xi\right]=0.
\label{eq:Dhalfparity}
\eeq
Adding these to $(\tilde\kappa_2^2+\chi_2^2)$ times \erf{eq:halfparityexpl} yields
\bqa
\Delta B_d(t)=\frac{-i \kappa_{12}}{2\chi_2}\tr \left[ \left(\big(\tilde\kappa_2-i\chi_2\Op\sigma_z^2-\frac{\tilde{\kappa_2}^2+{\chi_2}^2}{\kappa_2}\big)\Delta\Op\Pi_a\right.\right.\nn\\
 \left.\left.+(1-2\frac{\tilde\kappa_2}{\kappa_2})\Op{\dot\Pi}_a-\frac{1}{\kappa_2} \Op{\ddot \Pi}_a \right) \Op\Xi\right], \label{eq:dynamicBd}\nn\\
\eqa
with $\Delta\Op\Pi_a(t)=\Op\Pi_a(t)-\Op\Pi_a^{ad}(t)$.  \erf{eq:dynamicBd} fully specifies the shape parameterization of the compensation field $B_d(t)=B_d^{ad}(t)+\Delta B_d(t)$.  Note also that the derivative of $\Op\Pi_a$ is proportional to its deviation from the adiabatic value, $\Op{\dot\Pi}_a(t)=(\tilde\kappa_1-i\chi_1\Op\sigma_z^1)\Delta\Op\Pi_a(t)$.  Thus, in the bad cavity limit ($\kappa\gg \chi$), we find that the change in the compensation field relative to its adiabatic value is simply proportional to the deviation of the first cavity field from its adiabatic value. 

This solution for the optimal shaped pulse is demonstrated for an example of two cavities with identical frequencies but different transmittivities in Fig.~\ref{fig:pulseshaping}.  Here the solid blue line shows the pulse $A_d(t)$, the dotted orange line the compensation pulse $B_d(t)$ when the adiabatic approximation (\erf{eq:entangcond}) is used, while the dashed red line shows the exact solution of \erf{eq:dynamicBd}, which gives the optimal pulse shape that ensures the indistinguishability condition is met at all times. The parameters employed in this simulation are specified in section~\ref{subsec:trajectories}, with the only difference that we take unequal cavity transmittivities, $\frac{\kappa_1}{2\pi}$=3.9MHz and $\frac{\kappa_2}{2\pi}$=3.5MHz, for this calculation.  The exact compensation field is seen to be similar to the original pulse and the adiabatic approximation, with only slight changes during the transient periods of $A_d(t)$.  This ability to maintain the indistinguishability condition at all times by optimal shaping of the compensation pulse $B_d(t)$ is very relevant for experimental situations such as that in Ref.~\cite{Roch:2014ey}, where the qubits could be tuned to the same frequencies but the cavity losses are not identical.

\section{Results for lossy transmission}
\label{sec:results}
In this section we describe results from simulating the dynamics of the system in Fig. \ref{fig:apparatus} using the theoretical description developed in section \ref{sec:theory} together with experimentally realistic parameters. We use the polaron frame reduced master equation, \erf{eq:sme_polaron_red}, with a large number of random realizations of $dW(t)$ (60000 trajectories). This choice of frame is made here for computational efficiency, since we will only study observable values at the end of a measurement, when the cavities are unpopulated and hence the polaron and lab frames coincide. For these simulations we use the parameters $\frac{\chi_1}{2\pi}=1.2$MHz, $\frac{\chi_2}{2\pi}=1.0$MHz,  $\frac{\kappa_1}{2\pi}=18$MHz, $\frac{\kappa_2}{2\pi}=16$MHz, $\Delta_1=\Delta_2=0$, $\gamma_1= \frac{\kappa_1}{20}$,$\gamma_2= \frac{\kappa_2}{20} $, and $\sqrt{\gamma_1}A_d(t)=10$MHz. These parameters are representative of the parameter regime currently accessible in superconducting cavity-QED architectures \cite{Roch:2014ey}. The compensation field, $B_d(t)$, is chosen according to \erf{eq:entangcond} in order to ensure the distinguishability of the states in the single-excitation subspace, except during transients (\erf{eq:dynamicBd} was not used in this calculation, since as described above, this full compensation requires complex pulse shaping and hence is experimentally more challenging).

\begin{figure}
\centering \includegraphics[width=1.0\columnwidth]{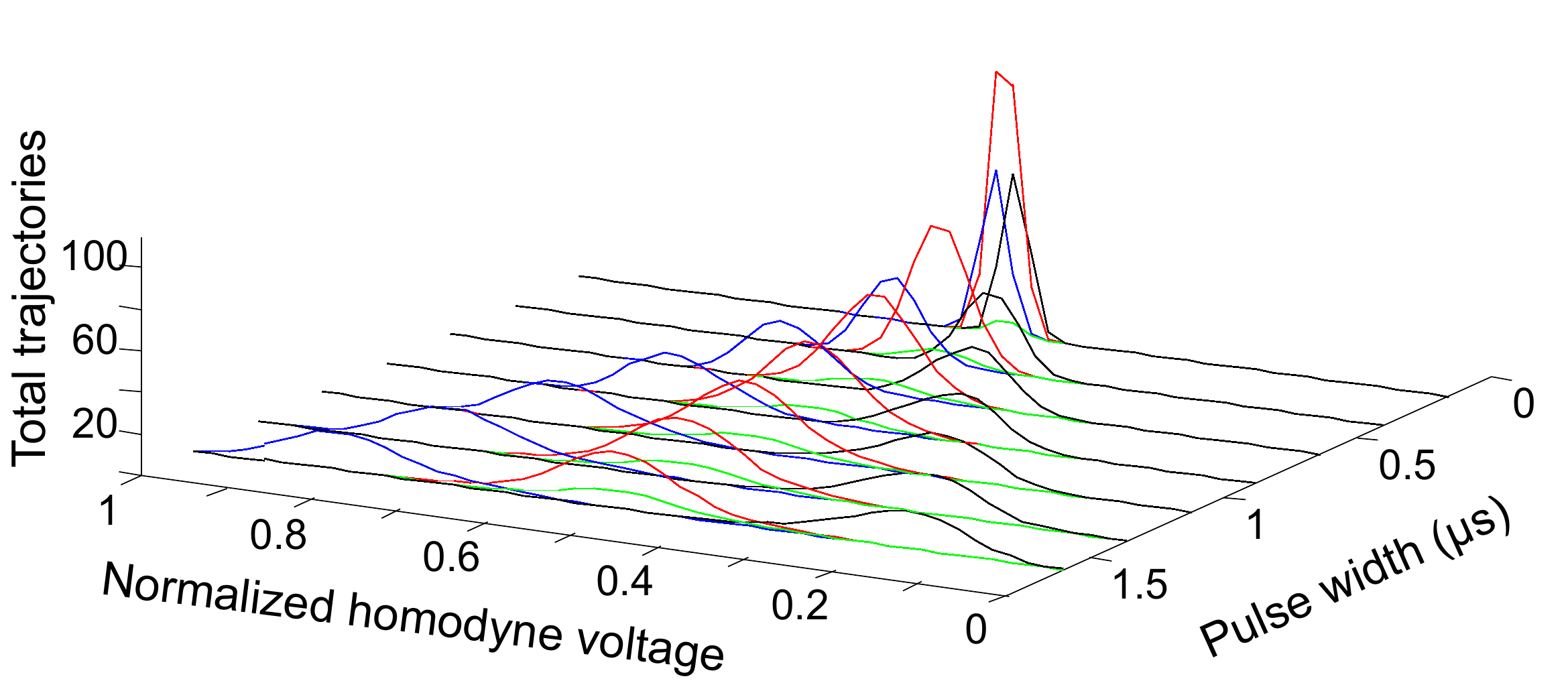}
\centering \includegraphics[width=0.9\columnwidth]{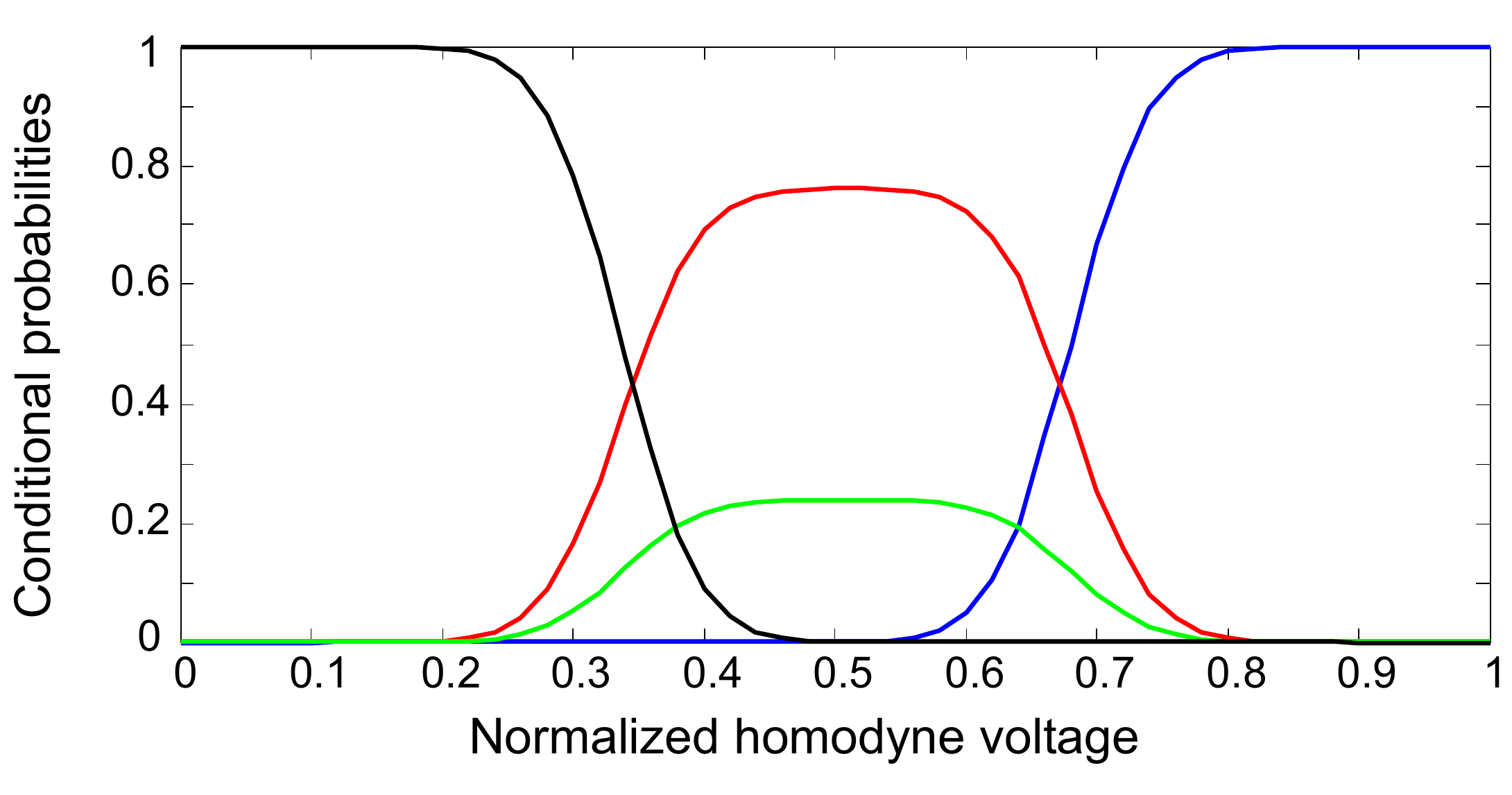}
\caption{(Color online) Distribution of qubit state populations (over 60000 trajectory simulations) as a function of normalized homodyne voltage and measurement pulse width. The initial state is given in \erf{eq:init}. Black (low voltage) represents $|00\rangle$ population, blue (high voltage) $|11\rangle$, red (upper, medium voltage) $\frac{1}{\sqrt{2}}(|01\rangle+|10\rangle )$, and green (lower, medium voltage) represents $\frac{1}{\sqrt{2}}(|01\rangle-|10\rangle )$.  The top panel plots relative frequencies of each of the state populations, while the bottom panel plots normalized populations conditioned on the measurement voltage value from the x-axis.  The bottom plot is essentially a slice of the top panel taken at a measurement pulse duration of 1$\mu$s. Simulation parameters are specified in section \ref{sec:results}. \label{fig:PvsTvsJ}}
\end{figure}

In a pulsed measurement setup the width of the measurement pulse dictates how well resolved the qubit states become. In Fig.~\ref{fig:PvsTvsJ} (top panel) we show how the state populations are distributed as a function of the (normalized) homodyne voltage and the measurement pulse width, over the 60000 simulated trajectories.  The bottom panel shows a slice at measurement pulse width of $1\mu$s, where the populations are normalized (such that the total population over these four orthogonal state is fixed to be one) at each homodyne voltage value. We see that for short pulse widths little information is carried out of the cavities and all qubit states are equally likely (low SNR). As a result of employing the compensation pulse $B_d(t)$ we see that, for all pulse widths, the $\ket{01}$ and $\ket{10}$ states are indistinguishable by the homodyne voltage value. In addition, for pulse widths greater than $\sim$ 1$\mu$s, we find that the homodyne voltages concentrate around the center value, predicting that primarily the single excitation subspace is populated under these conditions.  The presence of the $\frac{1}{\sqrt{2}}\ket{01}+\ket{10}$ state indicates that the indistinguishability condition has been satisfied between $\ket{01}$ and $\ket{10}$.  However, even if this is the case, other sources of dephasing could result in a mixed state with no entanglement for which the homodyne voltage would also be concentrated around the center value. For this reason we also plot the population of the antisymmetric state $\frac{1}{\sqrt{2}}(\ket{01}-\ket{10})$ (green line), presence of which would indicate a mixture with reduced or no entanglement, depending on the relative value of this state. We see that the antisymmetric state does not contribute for measurement pulse widths $\sim 1 \mu$s, but that its population increases as the measurement pulse width increases, as a result of intrinsic dephasing becoming significant at longer times. Fig.~\ref{fig:PvsTvsJ} therefore shows that there is a tradeoff between achieving high SNR with long measurement pulses and compensation for indistinguishability, and restricting the pulse duration to avoid intrinsic dephasing at longer timescales. 

To explore the influence of the two primary detrimental effects to achieving entanglement between qubits in separate cavities, namely loss between cavities ($1-\eta_l$) and measurement inefficiency ($1-\eta_{\rm m}$), we quantify the maximum achievable entanglement, quantified here by the concurrence \cite{Hil.Woo-1997}, for a range of these parameters in Fig. \ref{fig:contconc}. The maximal achievable concurrence (maximized over the measurement pulse width, over a range $0.1 - 4 \mu$s, and averaged over all trajectories) is plotted as a function of both loss of photons between the cavities ($\eta_l$ in dB) and measurement efficiency $\eta_{\rm m}$. In these simulations the indistinguishability criteria is enforced by the adiabatic compensation pulse, \erf{eq:entangcond}. The transmission losses between the two cavities lead to dephasing of the first qubit, and hence degrade the entangled state formed with the second qubit.  On the other hand, decreased measurement efficiency does not by itself lead to lower coherence, but instead necessitates longer measurement pulse widths or larger probe field amplitudes (\ie we need to use more photons to obtain the same amount of information). Nevertheless, low measurement efficiency combined with loss between cavities is detrimental, because even though one can increase the number of photons used to probe the qubits, this results in more photons being lost between cavities and thus in greater dephasing. Therefore the combination of low measurement efficiency together with large transmission loss between cavities is the most unfavorable situation for generating qubit entanglement. 

\begin{figure}
\centering \includegraphics[width=0.99\columnwidth]{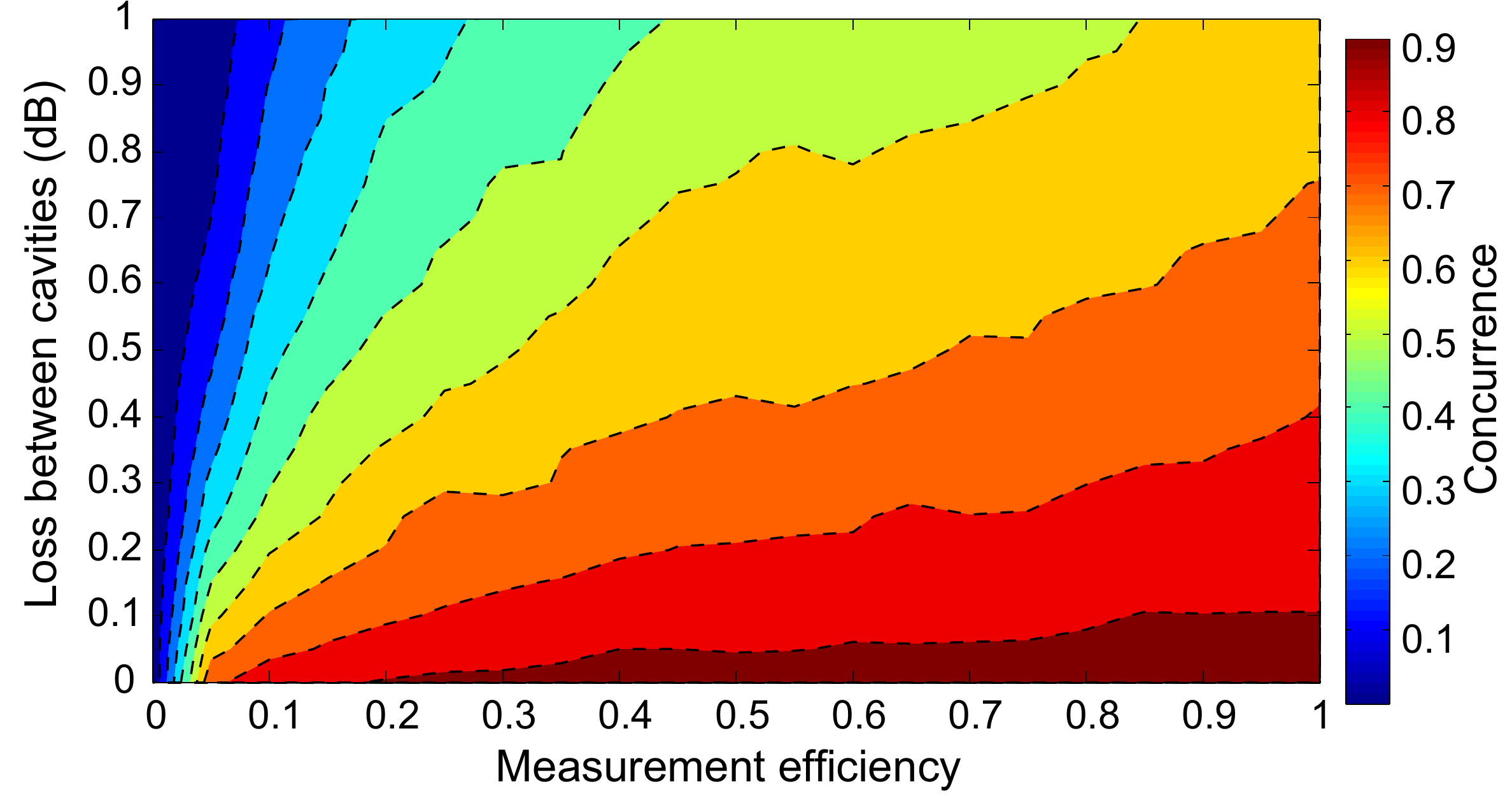}
\caption{(Color online) Maximum concurrence versus measurement efficiency ($\eta_{\rm m}$) and photon loss in the channel between the cavities ($1-\eta_l$, in dB).  Simulation parameters are specified in section \ref{sec:results}.  Entanglement maximum is calculated by propagating the simulation for different measurement pulse widths and selecting the maximal concurrence achieved for the given parameters averaged over all trajectories. }\label{fig:contconc}
\end{figure}

\section{Population transfer in the polaron frame}
\label{sec:CW}
The reduced model developed above and the analysis that followed are only valid when the original master equation, \erf{eq:full_me}, contains no terms that do not commute with the polaron transformation.  Two such terms that one might like to include in an extended model are qubit driving and population relaxation (T1 process). In this section we show how to perturbatively incorporate these effects into the qubit-only reduced master equations in \erf{eq:sme_polaron_red} or \erf{eq:qubit_sme}.

\subsection{Qubit driving}
Resonant drives on the qubits are described by addition of the Hamiltonian term
\beq
H_d=\Omega_1(t)\Op\sigma_x^{1}+\Omega_2(t)\Op\sigma_x^{2}.
\label{eq:driving}
\eeq
We will study instances where the magnitude of these new driving terms is small, and therefore they can be treated perturbatively.

As a result of the hybridization of the qubits and cavities in the polaron frame, qubit transitions will result in the cavity also being driven.  This can be seen from the effect of the two-cavity polaron transformation, \erf{eq:polaron} on the $\sigma_{-}^{i}$ operators:

\bqa
(\Op\sigma_-^{1})^P&=&D_2\left[\Op\Pi_b\right]D_1\left[\Op\Pi_a\right]\Op\sigma_-^{1} D\dg_1\left[\Op\Pi_a\right]D\dg_2\left[\Op\Pi_b\right]  \nn \\
(\Op\sigma_-^{2})^P&=&D_2\left[\Op\Pi_b\right]\Op\sigma_-^{2} D\dg_2\left[\Op\Pi_b\right]. 
\label{eq:polaron_sigma_minus}
\eqa
For simplicity, we consider an adiabatic probe pulse and compensation pulse. Then using the parameterization of \erf{eq:steadystates_general} we obtain to first order in $A_d\chi/\kappa^2$, the dressed $\sigma_{-}^{i}$ operators 
\bqa
(\Op\sigma_-^{1})^P&\approx&\Op\sigma_-^{1}(1 - A_d(\Op a+\zeta_2\Op b)\mu_1 + A_d^*(\Op a\dg+\zeta_2^*\Op b\dg)\mu_1^*),\nn \\
(\Op\sigma_-^{2})^P&\approx &\Op\sigma_-^{2}(1  -(B_d+A_d\zeta_1)\mu_2\Op b + (B_d^*+A_d^*\zeta_1^*)\mu_2^* \Op b\dg),\nn\\
\label{eq:axes}
\eqa
where $\mu_i=2\chi_i/((\kappa_i/2+i\Delta_i)^2+\chi_i^2)$ and $\zeta_i=\kappa_{12}(\kappa_i/2+i\Delta_i)/((\kappa_i/2+i\Delta_i)^2+\chi_i^2)$.

These dressed operators contain both qubit and cavity operators and thus in the polaron frame, qubit driving will also lead to cavity driving and damping. Since the cavities are detuned from the qubits the counter-rotating terms in the above expansion can be dropped in the RWA, and one obtains cavity sideband transitions involving the qubits (notably, driving the first qubit can result in a sideband transition in the second cavity due to the interconnection). Further, in the strong cavity damping limit, when $\Omega_i A_d\chi\ll \kappa_i^3$, the qubit driving Hamiltonian in the polaron frame, can be approximated as
\bqa
H_d^P &\approx & \Omega_1 \Op\sigma_x^{1}-\frac{{A_d}^2{\mu_1}^2(2{\Omega_1}^2 \Op\sigma_x^{1}+\Omega_1\chi_1\Op\sigma_z^{1})}{(\kappa_1/2+i\Delta_1 )^2-{\chi_1}^2-4 {\Omega_1} ^2}\nn\\
&+&\frac{{A_d}^2{\mu_1}^2{ \zeta_2} ^2(2{\Omega_1}^2 \Op\sigma_x^{1}+\Omega_1\chi_1\Op\sigma_z^{1})}{(\kappa_2/2+i\Delta_2 )^2-{\chi_1}^2-4 {\Omega_1} ^2}\nn\\
&+&\Omega_2 \Op\sigma_x^{2}-2{\lambda} ^2\frac{2{\Omega_2}^2 \Op\sigma_x^{2}+\Omega_2\chi_2\Op\sigma_z^{2}}{(\kappa_2/2+i\Delta_2 )^2-{\chi_2}^2-4 {\Omega_2} ^2},
\label{eq:driving_red}
\eqa
with ${\lambda}$=$B_d\mu_2+A_d\mu_2\zeta_1$. The cavity coupling induces a drive-dependent energy shift of the qubit, or alternatively, tilts the drive axis. This approximate drive Hamiltonian can be used with the qubit-only master equation, \erf{eq:sme_polaron_red}. However, effects outside of the qubit subspace will also remain in the polaron frame, such as sideband heating of the cavity.  It is evident from \erf{eq:driving_red} that the same control pulses will result in different angles of rotation, when viewed in the laboratory and polaron frames, and will only agree in the limit of zero photons in the cavities. In the dispersive limit and in steady state, the axes of rotation \erf{eq:axes} are constant and it is then straightforward to specify the $X$ and $Y$ rotation angles to achieve the desired rotation in either frame. 

In addition to this tilting of the qubit drive axis, the simultaneous application of a measurement tone and a qubit drive can lead to measurement-induced suppression of coherent oscillation (the Zeno effect), as noted for the single qubit case in Ref. \cite{Gam.Bla.etal-2008}.

\subsection{Qubit relaxation}
For long measurement pulses or continuous wave measurement, incorporating the effects of qubit relaxation ($T_1$ process) can become important. Similar to intrinsic dephasing, relaxation is incorporated into the full master equation, \erf{eq:full_me}, by the addition of the Lindblad terms
\beq
\mathcal{L}_r \varrho = \sum_{i=1}^2 \gamma_r^i \mathcal{D}[\Op \sigma_-^i]\varrho,
\label{eq:relax}
\eeq
where $\gamma_r^i$ are the intrinsic relaxation rates. The approximation \erf{eq:axes} to the polaron frame form for $\Op \sigma_-^i$ \erf{eq:polaron_sigma_minus} illustrates the problems that will arise when attempting to carry out the derivation of reduced master equations as in section \ref{sec:theory} in the presence of this relaxation term. Specifically, the presence of field excitation terms in the polaron frame mean that the cavity states are now no longer in the vacuum at all times in this frame. Physically, this reflects the fact that the relaxation process can create cavity photons which are not compensated by the polaron frame transformation. This means that \erf{eq:sme_polaron_red} is no longer a valid equation of motion for the qubit degrees of freedom in the polaron frame. Furthermore, \erf{eq:varrho_polaron_simple} is not a valid ansatz for the state of the qubit and cavity degrees of freedom. However, as we will show below, when the relaxation is slow and in the bad cavity limit, one can return to the lab frame and derive a valid reduced equation of motion in this frame that incorporates the qubit relaxation.

We begin with the more general representation of the state of the system in the polaron frame, that takes into account that the cavity states are no longer always the vaccum in this frame:
\beq
\varrho^P(t) = \hspace{-0.2cm}\sum_{\substack{n_an_bm_am_b \\ ijkl}} r_{n_an_bm_am_b,ijkl}(t) \ket{ij}\bra{kl} \otimes \ket{n_a n_b}\bra{m_a m_b}, \nn
\label{eq:varrho_polaron}
\eeq
where the indices $i,j,k,l$ run over $0$ and $1$, while the other indices (that index photon number states) run from $0$ to $\infty$.
Carrying out the same reduction to a representation of the state of the qubits in the lab frame made in section \ref{sec:reduced_lab} (\erf{eq:rho_qubits}), one finds that the mapping between the diagonal and off-diagonal terms in the polaron frame and the reduced lab frame is:
\begin{widetext}
\bqa
\rho_{ijij}(t) &=& \sum_{n_an_b} r_{n_an_bn_an_b,ijij}(t), \label{eq:pops_reln}\\
\rho_{ijkl}(t) &=& \sum_{n_an_bm_am_b} r_{n_an_bm_am_b,ijkl}(t) \bra{m_a m_b, kl} D_1\dg[\Op \Pi_a(t)]D_2\dg[\Op \Pi_b(t)]D_1[\Op \Pi_a(t)]D_2[\Op \Pi_b(t)]\ket{n_a n_b, ij} \label{eq:cohs_reln}\nn \\
&\equiv& \sum_{n_an_bm_am_b} \lambda_{n_an_bm_am_b,ijkl} (t).
\eqa
\end{widetext}
In the presence of qubit relaxation, calculating the time derivative of these off-diagonal elements is more involved since we cannot assume that $r_{0000,ijkl}$ is the only nonzero value in this expansion. 

Let us focus on the effect of the relaxation term \erf{eq:relax} alone. Incorporation of this term adds the following to the derivatives of the diagonal terms (ignoring all the other terms in the dynamical equation):
\bqa
\dot{r}_{n_an_bn_an_b,0000}(t) &=& \gamma_r^1 r_{n_an_bn_an_b,1010}(t) \nn \\
&& + \gamma_r^2 r_{n_an_bn_an_b,0101}(t) \nn \\
\dot{r}_{n_an_bn_an_b,0101}(t) &=& \gamma_r^1 r_{n_an_bn_an_b,1111}(t) \nn \\
&& - \gamma_r^2 r_{n_an_bn_an_b,0101}(t) \nn \\
\dot{r}_{n_an_bn_an_b,1010}(t) &=& -\gamma_r^1 r_{n_an_bn_an_b,1010}(t) \nn \\
&& + \gamma_r^2 r_{n_an_bn_an_b,1111}(t) \nn \\
\dot{r}_{n_an_bn_an_b,1111}(t) &=& -(\gamma_r^1+\gamma_r^2) r_{n_an_bn_an_b,1111}(t). \nn
\eqa
When the sum over $n_a$ and $n_b$ prescribed in \erf{eq:pops_reln} is performed, one gets equations of motion for $\rho_{ijij}(t)$ that are consistent with a qubit relaxation process in the reduced lab frame. However, the effect on the off-diagonal components is not as straightforward.  To illustrate this, we shall focus on a single off-diagonal element and calculate the contribution of the relaxation term $\gamma_r^1 \mathcal{D}[\Op \sigma_-^1]\varrho$ to the evolution of $\rho_{0001}(t)$. Explicitly, we obtain (again ignoring all the other terms in the dynamical equation),
\bqa
\dot{\rho}_{0001}(t) &=& \sum_{n_an_bm_am_b}\dot{\lambda}_{n_an_bm_am_b,0001}(t) \nn \\
 &=& \gamma_r^1 \sum_{n_an_bm_am_b} \lambda_{n_an_bm_am_b,1011}(t) \nn \\
 &=& \gamma_r^1 \sum_{n_an_bm_am_b} r_{n_an_bm_am_b, 1011} \delta_{n_a,m_a} \nn \\ &&~~~~~\bra{m_b} D_2\dg[B^{(11)}]D_2[B^{10}] \ket{n_b} \nn\\
 &=& \gamma_r^1 \sum_{n_an_bm_b} r_{n_an_bn_am_b, 1011} e^{i\mathrm{Im}\{B^{(11)*}B^{(10)}\}} \nn \\
 && ~~~~~\bra{m_b} D_2[B^{(10)}-B^{(11)}] \ket{n_b}. \nn
\eqa
The summation in this expression is difficult to perform exactly. However, we may simplify this by assuming the bad cavity limit and using the adiabatic values for the cavity fields given in \erf{eq:steadystates_general}. In this situation, 
\beq
|B^{(10)}-B^{(11)}| = \left| \frac{-i 2 \chi_2 B_d}{\tilde{\kappa}_2^2+\chi_2^2}\right| \ll 1,
\eeq
where the inequality in a consequence of the bad cavity limit. Similarly, in this limit we have $e^{i\mathrm{Im}\{B^{(11)*}B^{(10)}\}} \approx 1$. Now the matrix element $\bra{m_b}D_2[X] \ket{n_b}$ is proportional to the Laguerre polynomial $L^{n_b-m_b}_{n_b}(X)$, and is peaked around $n_a=n_b$ and zero everywhere else for small $X$ \cite{Wunsche:1991tx}. Therefore we approximate $\bra{m_b}D_2[B^{(10)}-B^{(11)}] \ket{n_b} \approx \delta_{n_b,m_b}$, and in this bad cavity limit (again for adiabatic values of the cavity fields), we find
\beq
\dot{\rho}_{0001}(t) \approx \gamma_r^1 \sum_{n_an_b} r_{n_an_bn_an_b,1011}(t) = \gamma_r^1 \rho_{1011}(t).
\eeq

Calculating the equations of motion for the other off-diagonal elements under the same approximations, we find that the total contribution of qubit relaxation to the lab frame reduced SME can be approximated by the addition of the following Lindblad term to \erf{eq:qubit_sme}
\beq
\mathcal{L}_r\rho(t) \equiv \sum_{i=1,2} \gamma_r^1 \mathcal{D}[\sigma_-^i]\rho(t),
\eeq 
and hence qubit relaxation carries through unaltered to the reduced SME in the lab dream.

\section{Summary and Conclusions}
\label{sec:disc}

We have derived a framework for describing joint dispersive measurement of qubits in separate cavities and shown how these measurements may be used to engineer entanglement between pairs of such qubits.  
The description shows how the populations of the qubit levels evolve diffusively as a function of a cascaded measurement and how calibrating the amplitude or frequency of compensation field(s) enables preservation of the coherence between the single-excitation states during the measurement.  We derived two alternative stochastic master equation approaches for the dynamical description of the qubit density matrix, one based on a polaron representation that describes dynamics in a dressed frame and the other a reduced master equation that described dynamics in the bare laboratory frame.  We derived static and dynamic entanglement conditions in the energy basis of the system, i.e., conditions on the physical parameters that guarantee the monitoring of the output field will ensure entanglement of the two qubits, and showed that these conditions give a simple prescription for ensuring this indistinguishability at all times.  We discussed simple  solutions of these conditions in three physically relevant limiting regimes of adiabatic probes, bad cavities, and an ideal setting of identical cavities with lossless transmission. We further showed how to achieve the indistinguishability condition at all times by optimal shaping of the compensation field $B_d(t)$.  The  theoretical analysis was applied to realistic experimental situations with extensive simulation of the different stochastic master equations in polaron and in laboratory frames.  The simulations show that the entanglement achieved with this procedure is tolerant to significant imperfections in the measurement efficiency and, to a lesser extent, to the presence of loss along the probe field path, \eg arising from circulators to ensure unidirectionality.  A detailed comparison was made between the simulations of the different qubit stochastic master equations in the polaron and laboratory frame.  A single trajectory analysis and related simulations revealed that the intra-cavity fields provide a non-Markovian environment resulting in suppression and revival of coherence of the bare laboratory qubit states, indicating that use of Markovian models for reduced qubit equations of motion in the lab frame is not always accurate.

The theoretical formulation presented here provides a first-principles description of the remote probabilistic entanglement achieved experimentally in Ref. \cite{Roch:2014ey}.  In addition, these results also motivate other schemes to achieve entanglement of superconducting qubits.  Of particular interest is the extension to continuous wave measurements rather than the pulsed measurements that were described here.  Continuous wave measurements are needed for longer and more complicated applications such as error correction protocols and feedback control in order to stabilize the entangled state in the presence of dephasing and relaxation processes. We have shown in this work that in the weak driving limit the main effect of simultaneous measurement and coherent control is a tilting of the axis of rotation that can be compensated (and a suppression of the rotation due to the measurement in the strong measurement limit, as was shown in Ref. \cite{Gam.Bla.etal-2008}). We have also shown that in the bad cavity limit phenomenological qubit relaxation terms simply carry over to the lab frame reduced master equation for the qubit degrees of freedom. This motivates introducing measurement-based feedback to create multi-qubit entanglement deterministically, which will be the topic of a future publication.

\begin{acknowledgements}
We gratefully acknowledge helpful conversations about the experimental details of the sequential probe apparatus with Nicholas Roch, Mollie Schwartz, R. Vijay, and Irfan Siddiqi.  We thank the Kavli Institute for
  Theoretical Physics for hospitality and for supporting this research in part by the National Science Foundation Grant No. PHY11-25915. FM and KBW were also supported in part by the National Science Foundation under the Catalzying International Collaborations program Grant
  No. OISE-1158954.
Sandia National Laboratories is a multiprogram laboratory managed and operated by Sandia Corporation, a wholly owned subsidiary of Lockheed Martin Corporation, for the United States Department of Energy's National Nuclear Security Administration under Contract No. DE-AC04-94AL85000.
\end{acknowledgements}

\appendix
\section{The SLH representation of quantum networks}
Building off work by Gardiner \cite{Gar-1993} and Carmichael \cite{Car-1993} on cascaded quantum optical systems, Gough and James have constructed a general formalism for modeling networks of quantum systems connected by bosonic fields. The utility of the approach is that it enables description of the dynamics of complex networks of modular components using simple composition rules. This formalism is primarily developed in Refs. \cite{Gou.Jam-2007, Gou.Jam-2009} and we summarize the main results in this Appendix.

The basis of the SLH modeling approach is to decompose a network into localized components with arbitrary degrees of freedom that are connected via freely propagating unidirectional broadband fields. This allows one to eliminate the fields propagating between components to arrive at an effective description of the system just in terms of the localized degrees of freedom and how they are connected together. 

The starting point for this formalism is the Hudson-Parthasarathy quantum stochastic differential equation (QSDE) accounting for time evolution of the unitary operator, $U(t)$, describing coupled evolution of the system and field degrees of freedom \cite{Hudson:1984wt}:
\bqa
dU(t) &=& \left\{ (S-1)d\Lambda(t) + L dB\dg(t) - L\dg S dB(t) \right. \nn \\
&& \left. - (\frac{1}{2}L\dg L + iH)dt\right\} U(t), 
\label{eq:hp_one}
\eqa
where $B(t)$ and $B\dg(t)$ are integrated versions of the freely propagating bosonic fields linearly interacting with the system at an interface or ``port" (these could be output fields from another system):
\bqa
B(t) = \int_0^t b(s) ds, ~~~~ B\dg(t) = \int_0^t b\dg(s) ds,
\eqa
with $[b(t), b\dg(s)] = \delta(t-s)$. This commutation relation defines the bosonic fields as rather singular objects, and hence the increments, $dB(t)=B(t+dt)-B(t)$ (and similarly of $dB\dg(t)$), are operator valued stochastic variables that are analogous to Ito increments. Finally, $\Lambda(t)$ is a quantum stochastic process that corresponds to the observable counting the number of quanta in the bosonic field that have interacted with the system up to time $t$:
\beq
\Lambda(t) = \int_0^t b\dg(s) b(s) ds
\eeq
The other components of \erf{eq:hp_one}, the system operators $S, L$, and $H$, describe the nature of the interaction between the system and propagating field at the interface. $S$ describes the impact on system when photons are scattered between ports (this component is most interesting when we consider systems with multiple ports, as we shall below), $L$ is the system operator that is directly and linearly coupled to the field, and $H$ is the system Hamiltonian that accounts for dynamics that does not involve interaction with the field $b(t)$. These components are often grouped together into a triple $G=(S,L,H)$, which is sufficient to completely characterize the system evolution. 

The generalization of \erf{eq:hp_one} to the case where the system has multiple ports, with independent fields at each port interacting with system, is:
\bqa
dU(t) &=& \left\{ \sum_{jk} (S_{jk}-\delta_{jk})d\Lambda_{jk}(t)  \right. \nn \\
&& + \sum_j L_j dB_j\dg(t) - \sum_{jk}L_j\dg S_{jk} dB_k(t) \nn \\
&& \left. - (\frac{1}{2}\sum_{j}L_j\dg L_j + iH)dt\right\} U(t), 
\eqa
where $S_{jk}$ describes the effect on the system of a photon scattering from port $j$ to $k$, and $L_j$ is the system operator coupled to the field at port $j$. In this multi-port case, we still describe this localized system with an SLH triple, but with the $S$ and $L$ now being matrix or vector valued: 
\bqa
S = \left(\begin{array}{ccc}S_{11} & \dots & S_{1n} \\ \vdots & \ddots & \vdots \\S_{n1} & \dots & S_{nn}\end{array}\right), ~~~~~ L = \left(\begin{array}{c}L_1 \\\vdots \\ L_n\end{array}\right)
\eqa
Note that the components of the $L$ vector are themselves operators. 

The key advantage of the SLH formalism is that one can easily construct effective descriptions of arbitrarily connected networks of localized components, each of which is represented by a triple: $G=(S,L,H)$. Connecting two components in series, parallel, or in feedback results in another system represented by another SLH triple whose matrices can be derived by simple algebraic rules \cite{Gou.Jam-2009}. For example, consider connecting two localized systems in series, where the outputs from $G_1=(S_1,L_1,H_1)$ and connected to the inputs of $G_2 = (S_2,L_2,H_2)$, where for simplicity we assume that the number of input ports that $G_2$ has is the same as the number of output ports that $G_1$ has. The resulting system is represented as 
\bqa
G_3 &=& G_2 \lhd G_1 \nn \\
&=& (S_3 \equiv S_2S_1, L_3 \equiv S_2L_1+L_2, \nn \\
&& ~~~~~ H_3 \equiv H_1+H_2+ \textrm{Im}\left\{L_2\dg S_2 L_1\right\}).
\eqa
Similarly, if one connects $G_1$ and $G_2$ in parallel (concatenates them), the resulting system is represented as
\bqa
G_3 &=& G_2 \boxplus G_1 \nn \\
&=& (S_3 \equiv \left(\begin{array}{cc}S_{1} & 0 \\ 0 & S_{2} \end{array}\right), L_3 \equiv \left(\begin{array}{c}L_{1} \\ L_{2} \end{array}\right), \nn \\
&& ~~~~~ H_3 \equiv H_1+H_2 ).
\eqa
See Refs. \cite{Gou.Jam-2009, Gough:2012fl} for more details on more complex composition rules.

\section{SLH triple for cascaded cavity apparatus}
The SLH triple that results from performing the concatenation and series products in \erf{eq:slh_prod} is
\begin{widetext}
\bqa
\Bigg(
\left[ \begin{matrix}
  \sqrt{\eta_l} & -i\sqrt{1-\eta_l} & 0 & 0 \\
  -i\sqrt{1-\eta_l} & \sqrt{\eta_l} & 0 & 0 \\
  0 & 0 & -1 & 0 \\
  0 & 0 & 0 & -1
\end{matrix}
 \right],
 \left[ 
 \begin{matrix}
  i\sqrt{1-\eta_l}(-\bar{\epsilon}(t) + \sqrt{\kappa_1} \Op a)  \\
  \sqrt{\eta_l}\bar{\epsilon}(t) -\sqrt{\kappa_1\eta_l} \Op a + \sqrt{\kappa_2}\Op b  \\
  -\bar{A}_d(t)+\sqrt{\gamma_1}\Op a  \\
  -\bar{B}_d(t) + \sqrt{\gamma_2}\Op b 
\end{matrix}
 \right], && \nn \\ 
 \Delta_{1} \Op a\dg \Op a + \chi_1 \Op a\dg \Op a \Op \sigma_z^1 + \Delta_{2} \Op b\dg \Op b + \chi_2 \Op b\dg \Op b \Op \sigma_z^2  + \frac{i}{2}(\Op a\dg A_d(t) - \Op a A_d^*(t)) + \frac{i}{2}(\Op b\dg B_d(t) &-& \Op b B_d^*(t)) - \frac{i\kappa_{12}}{2}(\Op a\dg \Op b - \Op b\dg \Op a))
 \Bigg)
\eqa
\end{widetext}

For any system represented by an SLH triple $(S,L,H)$, the corresponding master equation that describes the dynamics of the internal states in the model, represented by the density matrix $\rho$, is:
\beq
\ddt \rho = -i[H,\rho] + \sum_{k} L_k \rho L_k\dg - \frac{1}{2}L_k\dg L_k \rho - \frac{1}{2} \rho L_k\dg L_k,
\label{eq:gen_lindblad}
\eeq
where $L_k$ are the (operator-valued) elements of $L$. Notice that the scattering matrix $S$ does not influence the internal dynamics of the system, it only determines the relation between the input and output fields.

Finally, using the fact that the evolution of a density matrix under \erf{eq:gen_lindblad} is invariant under the transformations
\bqa
L_k &\rightarrow & L_k + \alpha \nn \\
H &\rightarrow & H - \frac{i}{2}(L_k \alpha^* - L_k\dg \alpha), \nn
\eqa
for $\alpha \in \mathbb{C}$, we can remove the terms proportional to identity in the $L$ vector to obtain the equivalent SLH triple
\begin{widetext}
\bqa
\Bigg(
\left[ \begin{matrix}
  \sqrt{\eta_l} & -i\sqrt{1-\eta_l} & 0 & 0 \\
  -i\sqrt{1-\eta_l} & \sqrt{\eta_l} & 0 & 0 \\
  0 & 0 & -1 & 0 \\
  0 & 0 & 0 & -1
\end{matrix}
 \right],
 \left[ 
 \begin{matrix}
  \sqrt{\kappa_1} \Op a)  \\
  -\sqrt{\kappa_1\eta_l} \Op a + \sqrt{\kappa_2}\Op b  \\
  \sqrt{\gamma_1}\Op a  \\
  \sqrt{\gamma_2}\Op b 
\end{matrix}
 \right], && \nn \\ 
 \Delta_{1} \Op a\dg \Op a + \chi_1 \Op a\dg \Op a \Op \sigma_z^1 + \Delta_{2} \Op b\dg \Op b + \chi_2 \Op b\dg \Op b \Op \sigma_z^2  + i(\Op a\dg A_d(t) - \Op a A_d^*(t)) + i(\Op b\dg B_d(t) &-& \Op b B_d^*(t)) - \frac{i\kappa_{12}}{2}(\Op a\dg \Op b - \Op b\dg \Op a))
 \Bigg)
\eqa
\end{widetext}
After addition to phenomenological dephasing terms with rate parameters $\gamma_d^i, i=1,2$, this SLH triple corresponds to the full master equation given in \erf{eq:full_me}. Each element of $L$ corresponds to an ``output" port that leaks photons, and the second port is the only one that is monitored (see Figs.  \ref{fig:apparatus} and \ref{fig:apparatus_slh}).

\bibliography{entanglement_longpaper}

\end{document}